\documentclass[pra,superscriptaddress,aps, bibliography, twocolumn, reprint, showpacs, footnoteinbib]{revtex4-1}

\usepackage{enumerate}
\usepackage{braket}  
\usepackage{graphicx}
\usepackage{amssymb} 
\usepackage{amsfonts}
\usepackage{amsmath}
\usepackage{dsfont}
\usepackage{natbib}
\usepackage{dcolumn} 
\usepackage{bm}  
\usepackage[mathscr]{eucal}
\usepackage[dvipsnames]{xcolor}
\usepackage[colorlinks, linkcolor=OrangeRed,citecolor=RoyalBlue,urlcolor=NavyBlue]{hyperref}
\usepackage[all]{hypcap} 
\usepackage{mathtools}
\usepackage{wasysym}

\newcommand{\mc}{\mathcal}
\newcommand{\ovl}{\overline}

\newcommand{\rev}[1]{{\color{black}#1}}

\usepackage{soul}

\begin{document}

\title{Return probability for the Anderson model on the random regular graph}
\author{Soumya Bera}
\affiliation{Department of Physics, Indian Institute of Technology Bombay, Mumbai 400076, India}
\author{Giuseppe De Tomasi}
\email[]{detomasi@pks.mpg.de}
\affiliation{Max-Planck-Institut f\"ur Physik komplexer Systeme, N\"othnitzer Stra{\ss}e 38, 01187-Dresden, Germany}
\author{Ivan M. Khaymovich}
\affiliation{Max-Planck-Institut f\"ur Physik komplexer Systeme, N\"othnitzer Stra{\ss}e 38, 01187-Dresden, Germany}
\author{Antonello Scardicchio}
\affiliation{Abdus Salam International Center for Theoretical Physics, Strada Costiera 11, 34151 Trieste, Italy}
\affiliation{INFN, Sezione di Trieste, Via Valerio 2, 34126, Trieste, Italy}
\begin{abstract}
\rev{We study the return probability for the Anderson model on the random regular graph and give the evidence of the existence of two
distinct phases: a fully ergodic and non-ergodic one.
In the ergodic phase the return probability decays polynomially with time with oscillations, being the attribute of the Wigner-Dyson-like behavior, while in the non-ergodic phase the decay follows a stretched exponential decay.
We give a phenomenological interpretation of the stretched exponential decay in terms of a classical random walker.
Furthermore, comparing typical and mean values of the return probability, we show how to differentiate an ergodic phase from
a non-ergodic one. We benchmark this method first in two random matrix models, the power-law random banded matrices and the Rosenzweig-Porter matrices, which host both phases. Second, we apply this method to the Anderson model on the random regular graph to give further evidence of the existence of the two phases.}
\end{abstract}
\maketitle
\section{Introduction}
The problem of Anderson localization on locally tree-like structures~\cite{Abou73,mirlin1991localization}, or Bethe lattices, which are limits of families of random regular graphs (RRG), has been at the center of a recent spur of research activity~\cite{Deluca14,Alt16,Alts16,Tik16,TikMir16,garcia2017scaling,Sonner17,Biro17,Kra18}. A big role in this renaissance has been played by the connections of this problem with that of localization in interacting quantum systems, dubbed many-body localization (MBL)~\cite{Basko06}. In fact, the original idea of mapping a disordered quantum dot to a localization problem in a section of the Fock space~\cite{Alt97} has been a quite useful paradigm to follow, on the route to a more accurate description of localized, interacting systems used by Basko, Aleiner and Altshuler in their seminal work~\cite{Basko06}.

The MBL phase~\cite{nandkishore2015many}, which now looks like the prototypical dynamical behavior of an interacting quantum system with strong disorder, has been characterized completely in terms of emergent,
local integrals of motion~\cite{huse2013phenomenology,serbyn2013local,ros2015integrals,Chandran2016many,Nandkishore2016emergent} (for a review see~\cite{abanin2017recent,imbrie2017review}).
However, a similar degree of understanding of the phase transition or of the delocalized region at smaller disorder is lacking (see~\cite{luitz2017ergodic} for a recent review). Many numerical works analyzing spin chains in one dimension did indeed confirm the MBL transition~\cite{Pal10, de2013ergodicity, Luitz15, Rajeev16,Giu17, Bera15, Ser15,pietracaprina2017entanglement}, but critical exponents are in disagreement with very general bounds~\cite{chandran2015finite}, hence casting doubts on the fact that the system sizes analyzed are in the scaling region.

Other works have found a diffusive to subdiffusive phase transition in the delocalized region~\cite{Luitz16,Scardi16,La16}
(but~\cite{Kara09,Ste16,Bera17} questioned these findings). Subdiffusion has been interpreted sometimes \rev{in terms of} rare-region effects~\cite{La16,Luitz16,Reich15,Karr17,Khait16,Agar15,Vosk15,Gopa16,Zhang16}, but this interpretation should be questioned as it is found also in models with quasi-periodic disorder in which Griffith effects are suppressed.

In light of these findings, and if the mapping of the MBL problem to the Anderson model on the RRG has to be taken to its extreme consequences, one is led to wonder if different
\emph{flavors} of the delocalized phase should be present there too (this is at some level conjectured in~\cite{Alt97}). This is an intriguing possibility, and an interesting question \emph{per se}.
Since numerical analysis of the Anderson model on $\mathbb{Z}^d$ lattices for small $d$ (mainly up to $d=6$~\cite{Evers2008Review,PhysRevB.95.094204}) found no such phase, this possibility is clearly linked to the nature of the RRG,
or to mean field approximations valid when $d\to\infty$.
\rev{However, recently in long-range random matrix models such behavior has been found in several models \cite{Kravtsov_NJP2015,Nosov2018correlation,Khaymovich_misc}.
More or less simultaneously,} it has been proposed that the Anderson model on the RRG might have a new phase within the extended phase (where states span the entire space). In this non-ergodic, extended (NEE) phase, multifractal states \rev{possess} strong fluctuations in space~\cite{Deluca14,Alt16,Kra18, Alts16}, like the states exactly at the Anderson transition $W_{AT}$ do in the Anderson model on $\mathbb{Z}^d$. Focusing on properties of the eigenfunctions, several studies propose numerical and analytical insights for the existence of this phase, but, lacking an exact solution, the existence of this NEE phase has been strongly questioned giving an indication that it could be just a finite size effect~\cite{Tik16,Sonner17,TikMir16,garcia2017scaling}, and the topic is generating an active debate.

In this work, we focus on the characterization of the delocalized phase, basing on time evolution of observables, \rev{which more sensitive to non-ergodicity and converge at available system sizes, unlike} 
eigenfunction statistics. Studying the return probability of a particle initially localized in a small region of the system, we show how to spot the existence of multifractal states, emphasizing the importance of the fluctuations of the return probability. At the level of numerical simulations, typically, studying dynamics is easier than studying eigenfunctions, and one can reach larger systems sizes. In this paper we study the Anderson model (AM) on the RRG for $N$ up to $2^{20}\simeq 10^6$ vertices, while typically eigenfunction statistics is available up to $2^{17}$. We also show how the dynamical properties we focus on have converged at these sizes, while eigenfunctions observables, like the inverse participation ratio (IPR), have not.

First, we benchmarked this characterization on two known models that possess critical states: The power-law random banded matrix
(PLRBM)~\cite{Mirlin96,Evers2008Review}, and the Rosenzweig-Porter random matrix (RPRM) models~\cite{Rose60,kravtsov2015random}.
\rev{The former model mimics the Anderson localization transition at finite $d$, showing only ergodic and localized phases and giving the access to multifractal states at the Anderson transition point.
The latter exhibits an entire fractal phase~\cite{kravtsov2015random} in a range of parameters along with the standard ergodic and localized phases~\footnote{Recently, some works report on other models showing the presence of the non-ergodic extended phases
away from criticality (see, e.g., \cite{Yuzbashyan_JPhysA2009_Exact_solution,Yuzbashyan_NJP2016,Nosov2018correlation,Khaymovich_misc}).}.}

Then, we use the same concept to study the Anderson model on the RRG, \rev{showing} similarities and differences with the previous two models\rev{, PLRBM and RPRM.}
We find that the ratio of logarithms of the mean and typical values of the return probability is, to a good approximation, a constant.
While this constant \rev{equals to unity}
in the ergodic phase of the PLRBM and for the RPRM model, it is smaller than \rev{unity} for the multifractal phase of the PLRBM and
for the AM on the RRG.
\rev{In the ergodic phase in all models, both mean and typical values of the return probability show an universal algebraic decay with time with oscillations, due to the rigidity of their spectrum~\cite{deTomasi_2018,tavora2017power,torres2018generic,santos2018nonequilibrium}.}
\rev{This kind of ergodicity is usually referred to as full ergodicity and characterized by standard Gaussian ensembles with ergodic fully correlated wavefunctions and Wigner-Dyson level statistics~\cite{Mehta}.}

\rev{In the multifractal phase of the PLRBM},  mean and typical values of return probability  are power-law \rev{decaying but} with different powers, \rev{while in the RPRM the decay is exponential.
In the multifractal phase of the AM on the RRG, the mean and typical values are stretched exponentials $R(t)\sim \exp[-\Gamma t^\beta]$}, with the same power $\beta$ but different pre-factors $\Gamma$.
Notice that the return probability for the AM on the RRG has been numerically investigated in~\cite{Biro17},
but for smaller system sizes and time \rev{scales}, where the stretched exponential was approximated with a power law.

\rev{Our analysis} gives a characterization of the delocalized region of the AM on the RRG, $W<W_{AT}$\rev{.
Indeed, at small enough disorder strengths, $W/W_{AT}<0.16$,
a fully ergodic phase is established, while for $W/W_{AT}$ from $0.4$ to $0.7$
a non-ergodic extended phase appears,
which is somehow intermediate between the PLRBM one and the fractal region of the RPRM.}
Moreover, within the range of values of disorder in which the non-ergodic phase has been found ($0.4 \le W/W_{AT} \le 0.7$),
\rev{the parameters of the stretched exponential $\beta$ and $\Gamma$ evolve smoothly with $W$, where $\beta\rightarrow 0$ as $W$ approaching the critical value  $W_{AT}$}.

In the last section we provide a classical random walk model in which the particle jumps in random directions but at random times $\Delta t$,
which are distributed in a power-law way $P(\Delta t)\sim (\Delta t)^{-(1+\beta)}$. The exponent $\beta\simeq 1-W/W_{AT}$ is the exponent of the stretched exponential.

\section{Model and methods}
We study the Hamiltonian
\begin{equation}
 \hat{H}:= \sum_{x,y=1}^{L} h_{x,y}|x \rangle \langle y|,
\label{eq:Ham}
\end{equation}
represented in the basis of the site states $|x\rangle$, where $L$ is the number of sites in the system.
We consider three different models that have a metal-insulator transition (MIT) with wave-functions changing properties
from ergodic to localized via multifractal ones.

First, we consider the power-law random banded matrix ensemble (PLRBM)~\cite{Mirlin96,Evers2008Review}, which is obtained from $\hat{H}$~(\ref{eq:Ham}) with $h_{x,y} = h_{y,x} = \mu_{x,y}/(1+(|x-y|/b)^{2a})^{1/2}$.
Here and further $\mu_{x,y}$ are independent uniformly distributed random variables taken from $[-1,1]$.
This ensemble of matrices parameterized by $a$ and $b$ \rev{has} an MIT at $a=1$, for any $b$.
For $a<1$, the model shows an ergodic phase~\rev{\footnote{However, some recent works claim that the fully ergodic Wigner-Dyson wavefunction distribution realizes only at $a<1/2$, showing at $1/2<a<1$ weakly non-ergodic behavior though with ergodic wave function moments
(see, e.g.,~\cite{BogomolnyPLRBM2018,Nosov2018correlation}).}} and at $a>1$ the eigenstates are power-law localized.
At the critical point ($a=1$) all the states are multifractal and the parameter $b$ tunes the
multifractal properties of eigenstates from strong ($b\ll 1$) to weak ($b\gg 1$) multifractality~\cite{Mirlin96,Mirlin00,Levitov90}.
There is no mobility edge in this model, i.e. for any $a,~b$ all the states are either extended or localized~\footnote{Note that the model with deterministic hopping terms $h_{x,y}=1/(1+(|x-y|/b)^{2a})^{1/2}$ at $x\ne y$, considered first in~\cite{Burin1989}, shows always power-law localized states for any $a$~\cite{Deng2018duality,Nosov2018correlation}.}.

Second, \rev{we discuss} the Rosenzweig-Porter random matrix model (RPRM)~\cite{Rose60,Kravtsov_NJP2015}\rev{, which} is obtained choosing $h_{x,y}=h_{y,x} = \mu_{x,y}/L^{\gamma/2}$ for $x\ne y$, while for $x=y$, $h_{x,x} = \mu_{x,x}$.
Like the PLRBM the RPRM has no mobility edge, but it has three distinct phases. For $\gamma<1$ all the states are \rev{fully} ergodic
while at $\gamma>2$ all the states are localized \rev{nearly at a single site~\cite{Kravtsov_NJP2015,deTomasi_2018}}. For $1<\gamma <2$ a simple fractal phase arises, (one does not have multifractality)~\footnote{Note that the model with deterministic hoppings $h_{x,y} = g_x g_y L^{-\gamma/2}$ with fixed $g_x\sim L^0$ being an integrable model is either localized or critical~\cite{Owusu2008link,Modak2016integrals,Borgonovi_2016,Deng2018duality,Khaymovich_misc,Nosov2018correlation}.}.
In a fractal phase of the RPRM, the wave-functions can  be  considered  ergodic on a large number of sites (\rev{which form a fractal}), which is however a small
fraction of the whole system \rev{(zero measure in the thermodynamic system $L\to\infty$)}.
The consequence of this, is that the exponents $\tau_q$ of a certain eigenstate $\phi_E(x)$ of the Hamiltonian $\hat{H}$, defined by $\sum_x |\phi_E(x)|^{2q}\sim L^{-\tau_q}$ \rev{take a simple linear} form $\tau_q = (2-\gamma)(q-1)$, $q>1/2$.

Third, \rev{we examine} the random regular graph (RRG) \rev{with the uncorrelated diagonal disorder $h_{x,x}$ uniformly distributed in the interval} $[-W/2,W/2]$.
\rev{The hopping amplitudes are deterministic and equal to} $h_{x,y} = h_{y,x} = -1$ if the sites $x$ and $y$ are linked in RRG with fixed local connectivity \rev{$K+1$} and $h_{x,y} =0$ otherwise.
The local connectivity is taken to be three \rev{(i.e., $K=2$)} like in \rev{many} previous studies.
This model is believed to have the Anderson transition (AT) at $W_{AT} \approx 18.2$
(this number is the most recent one in~\cite{Kra18} and~\cite{Parisi_misc}).
Moreover, \rev{the matter of discussion is the possibility of the existence of an
non-ergodic (multifractal) phase constituted} by extended states at $W<W_{AT}$~\cite{Deluca14} and \rev{thus a}
transition at even smaller disorder strength between these multifractal states and ergodic states~\cite{Alt16}.
This putative transition has been estimate to be around $W_{EMT} \approx 10$ (EMT, ergodic to multifractal)~\cite{Alt16}.
It implies existence of an entire phase ($W_{EMT}<W<W_{AT}$) composed of multifractal states.
The RRG has mobility edges, thus the spectrum of $\hat{H}$ depending on the disorder strength can host separated bands of energies composed of extended or localized eigenstates.

\rev{In this work we focus on the study of }
these different extended phases (ergodic, non-ergodic multifractal, fractal) by investigating
their dynamical properties. In particular, we study the return probability starting from a \rev{projected} state \rev{$\hat{P}_{\Delta E}|x\rangle$}~\cite{deTomasi_2018,Bera17,detomasi16}, defined as:
\begin{equation}
 \mc{R}(t) := \frac{|\langle x | \hat{P}_{\Delta E} e^{-i \hat{H} t } \hat{P}_{\Delta E} |x \rangle |^2}{
|\langle x | \hat{P}_{\Delta E} |x \rangle |^2},
\end{equation}
where \rev{for RRG} $\hat{P}_{\Delta E} := \sum_{E\in \Delta E} | E\rangle \langle E|$ is the projector to eigenstates of $\hat{H}$ with energy $E$
which belongs to a small energy shell $E\in \Delta E = [-\delta E, \delta E]$ around the middle of the spectrum of $\hat{H}$, $|\Delta E| = 2\delta E=E_{BW}/32$ is considered to be a fraction of the whole bandwidth $E_{BW}$ for the Anderson model on the RRG. \rev{For PLRBM and RPRM (where there is no mobility edge) the projector is taken to be} $\hat{P}=\mathbb{I}$. 

The reason to use the projector $\hat{P}_{\Delta E}$ in the Anderson model on the RRG is dual. On one hand one wants to avoid the mixing of states with different dynamical properties, and in general some $\ket{x}$ have overlap with both localized and delocalized states. So, for the RRG, $\Delta E$ has been chosen small enough so that the eigenstates involved in the dynamics are almost all extended for the values of considered disorder~\footnote{We tested that our result barely depend on 
the choice of the fraction $|\Delta E|/E_{BW}<1/8$. Please see~Appendix for the data with $|\Delta E|/E_{BW}$ from $1/64$ to $1/8$.}. On the other hand, one would like to create a \emph{semiclassical} wave packet, in hope that some kind of classical random process can capture the quantum dynamics. So one has to balance of the \rev{uncertainties} $\Delta x$ and $|\Delta E|$ such that the \rev{uncertainty} principle $\Delta x|\Delta E|\gtrsim \hbar v$ is satisfied (here $v$ is some velocity $O(1)$).

The average over matrix ensemble and initial states $|x\rangle$ is indicated with a \rev{bar} over the quantity considered.
In particular, we \rev{focus on} mean and typical values of $\mc{R}(t)$, defined as $\ovl{\mc{R}}(t)$ and
$e^{\ovl{\log{\mc{R}(t)}}}$\rev{, respectively}.

The scaling of $\mc{R}(t)$ to zero with the system size $L$ in the long time limit is also in our main focus (both typical and mean averages)
\begin{equation}
 \ovl{\mc{R}_{\infty}} := \lim_{T\rightarrow\infty} \frac{1}{T} \int_{0}^T \ovl{\mc{R}}(t) dt,
 \label{eq:mean_long}
\end{equation}

\begin{equation}
 e^{\ovl{\log{\mc{R}_{\infty}}}} := \lim_{T\rightarrow\infty} \frac{1}{T} \int_{0}^T e^{\ovl{\log\mc{R}}(t)}dt.
 \label{eq:typical_long}
\end{equation}

These quantities will give information on the properties (ergodicity or multifractality) of the eigenstate belonging in the energy shell $\Delta
E$ as the mean $\ovl{\mc{R}_\infty}$ can be expressed in terms of the inverse participation ratio (IPR) of wavefunctions $\{\phi_E\}$ of $\hat{H}$
\footnote{Note that the considered definition of the inverse participation ratio, $IPR_x$, is different from the standard one $IPR_E = \sum_{x} |\phi_E(x)|^4$ as the summation is taken over energy window, but not over sites.},
\begin{equation}
\ovl{\mc{R}_\infty} = \ovl{IPR_x} = \ovl{\frac{\sum_{E\in \Delta E} |\phi_E(x)|^4}{ (\sum_{E\in \Delta E} |\phi_E(x)|^2)^2}}.
\label{eq:relation}
\end{equation}
The typical value $e^{\ovl{\log{IPR_x}}}$ of $IPR_x$ is not equal to
$e^{\ovl{\log{\mc{R}_{\infty}}}}$ in general. This difference \rev{possibly originates from} the time fluctuations of $\mc{R}(t)$.
Nevertheless for long times (of the order of the saturation \rev{time of $\mc{R}(t)$ in} a finite system)
the time fluctuations of $\mc{R}(t)$ scale to zero as a function of $L$, so in the first approximation
the correction due to time fluctuations does not change the $L$-scaling of $e^{\ovl{\log{IPR_x}}}\sim e^{\ovl{\log{\mc{R}_{\infty}}}}$.
We confirmed the scaling $e^{\ovl{\log{IPR_x}}}\sim e^{\ovl{\log{\mc{R}_{\infty}}}}$ with exact numerics.

Nevertheless, the scaling of $\ovl{IPR_x}$ and $e^{\ovl{\log{IPR_x}}}$ can, in principle, be different depending on the phase.
Indeed, in the ergodic phase the envelope of the wavefunctions $\{\phi_E\}$ is in the first approximation uniformly distributed over the entire system $(|\phi_E(x)|^2\sim 1/L)$,
thus it does not reveal strong spatial fluctuations. In this case, we do not expect any difference in the scaling of mean and typical values.
In a fractal phase, like in the RPRM, the magnitude of wavefunctions in space do not \rev{possess significant} fluctuations, since the fractality is
\rev{emerged due to a fractal spatial support set of wavefunctions, forming
subbands in the entire energy spectrum from eigenstates living in the same fractal set and fully correlated to each other~\cite{Kravtsov_NJP2015,deTomasi_2018}. Thus, in this case} we expect a situation similar to that of the ergodic phase.
Nevertheless, in the multifractal phase the wave-functions $\{\phi_E(x)\}$ could have strong spatial dependence,
which could imply a possible difference in scaling with $L$ between $\ovl{\mc{R}_\infty}$ and $e^{\ovl{\log{\mc{R}_\infty}}}$.

 \begin{figure}[t]
 \includegraphics[width=0.95\columnwidth]{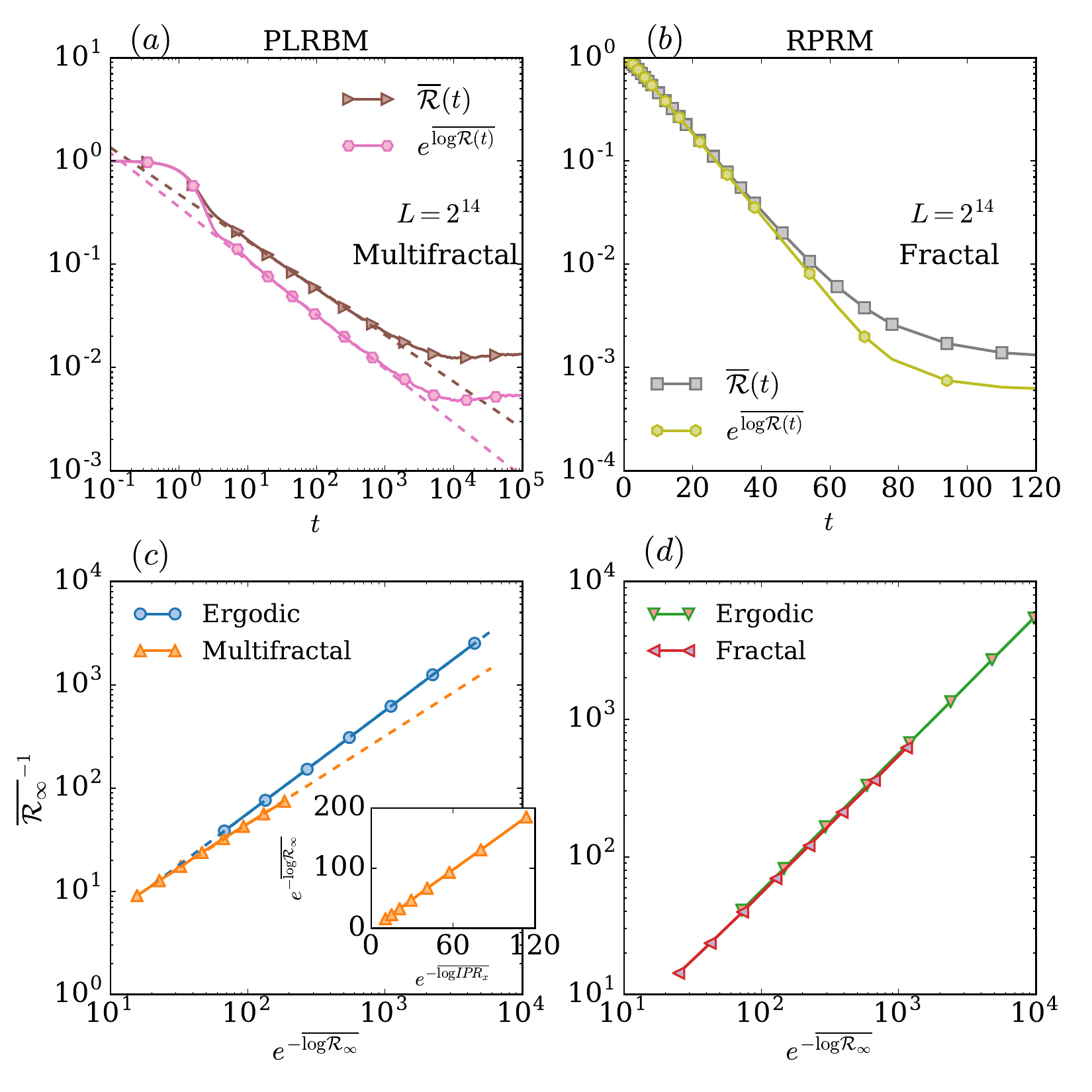}
 \caption{\rev{{\bf Mean $\ovl{\mc{R}}(t)$ and typical $e^{\ovl{\log{\mc{R}}}(t)}$ return probability as a function of time  $t$ (a, b) or versus each other (c, d) for PLRBM and RPRM models.}
 (a)~$\ovl{\mc{R}}(t)$ and $e^{\ovl{\log{\mc{R}}}(t)}$ versus $t$ shown in log-log scale in the multifractal phase of the PLRBM
 ($a=1,b=0.5$) for a fixed system size $L=2^{14}$.
 The dashed lines are guides for the eyes emphasizing the power-law decay
 of $\ovl{\mc{R}}(t)$ and $e^{\ovl{\log{\mc{R}}}(t)}$ in time with different powers.
 (b)~$\ovl{\mc{R}}(t)$ and $e^{\ovl{\log{\mc{R}}}(t)}$ versus $t$ shown in log-linear scale in the fractal phase of RPRM ($\gamma=1.25$) for $L=2^{14}$. $\ovl{\mc{R}}(t)\sim e^{\ovl{\log{\mc{R}}}(t)} \sim e^{-E_{Th}t}$ show exponential decay with the same exponent $E_{Th}$ being the Thouless energy of the model ($E_{Th}\sim L^{1-\gamma}$).
 (c)~$\ovl{\mc{R}_\infty}^{-1}$ versus $e^{-\ovl{\log{\mc{R}_\infty}}}$
 in two different phases of the PLRBM: In the ergodic phase ($a=0.5, b=1$) the mutual dependence is linear $\ovl{\mc{R}_\infty}^{-1}\sim e^{-\ovl{\log{\mc{R}_\infty}}}$,
 while in the multifractal phase ($a=1,b=0.5$) $\ovl{\mc{R}_\infty}^{-1}\sim e^{-\alpha \ovl{\log{\mc{R}_\infty}}}$ with $\alpha < 1$.
 The inset shows the scaling of the typical $IPR_x$ versus the typical $\mc{R}(\infty)$ in the linear scale and confirms the same $L$-scaling of both typical quantities.
 (d)~$\ovl{\mc{R}_\infty}^{-1}$ versus $e^{-\ovl{\log{\mc{R}_\infty}}}$ for the RPRM in two phases.
 Both in ergodic ($\gamma = 0.5$) and fractal ($\gamma = 1.25$) phases the mutual dependence is linear $\ovl{\mc{R}_\infty}^{-1}\sim e^{-\ovl{\log{\mc{R}_\infty}}}$.}
 }
 \label{fig:fig1_RP+PLBM}
 \end{figure}

\section{PLRBM $\&$ RPRM}
In this section we study $\mc{R}(t)$ and its long time saturation value for the PLRBM and RPRM.
We perform the time evolution using exact full diagonalization.
At the critical point of the PLRBM, $a=1$, where all states are multifractal, both $\ovl{\mc{R}}(t)$ and
$e^{\ovl{\log{\mc{R}(t)}}}$ decay algebraically, $\ovl{\mc{R}}(t)\sim t^{-\alpha_1}$
and $e^{\ovl{\log{\mc{R}(t)}}}\sim t^{-\alpha_2}$ in full agreement with the previous analytical investigations for $\ovl{\mc{R}}(t)$~\cite{kravtsov2010dynamical,kravtsov2011return,kravtsov2012levy}.
Figure~\ref{fig:fig1_RP+PLBM}(a) shows the algebraic decay of $\ovl{\mc{R}}(t)$
and $e^{\ovl{\log{\mc{R}(t)}}}$ at criticality (multifractal phase).
As observed, the two decay rates ($\alpha_1,\alpha_2$) are different from
each other, and due to the inequality between arithmetic and geometric mean $\alpha_1<\alpha_2$ \footnote{We analyzed also the ultrametric random matrix model, a different model in which a MIT happens. At its critical point all eigenstates are multifractal and the same difference
between mean and typical value of $\mc{R}(t)$ like in PLRBM holds}.
Instead, in the ergodic phase ($a<1$) the asymptotic decay rates of
$\ovl{\mc{R}}(t)$ and $e^{\ovl{\log{\mc{R}(t)}}}$ are the same
\rev{and $\mc{R}(t)$ demonstrates power-law decay with oscillations, being an attribute of Wigner-Dyson fully ergodic behavior}~(see Appendix).
As a consequence of \rev{the difference in decay rates of} $\ovl{\mc{R}}(t)$ and $e^{\ovl{\log{\mc{R}(t)}}}$ in the multifractal phase
the saturation values $\ovl{\mc{R}_{\infty}}$~(\ref{eq:mean_long}) and $e^{\ovl{\log{\mc{R}_{\infty}}}}$
(\ref{eq:typical_long}) may have different scaling to zero as functions of $L$, $\ovl{\mc{R}_{\infty}} \sim L^{-D_2}$
and $e^{\ovl{\log{\mc{R}_{\infty}}}} \sim L^{-D_{\text{typ}}}$ ($D_2<D_{\text{typ}}<1$),
while in \rev{the ergodic phase the exponents are the same and equal to unity} $D_2 = D_{\text{typ}} = 1$.

\rev{To emphasize the difference in the behavior of the typical and mean $\mc{R}(t)$ in different phases of PLRBM, in
Fig.~\ref{fig:fig1_RP+PLBM}(c) we show} $\ovl{\mc{R}_\infty}^{-1}$ as a function of $e^{-\ovl{\log{\mc{R}_\infty}}}$ in a log-log plot for two
different \rev{set of} values of $a,~b$: one in the ergodic phase and another in the multifractal phase.
In the ergodic phase $\ovl{\mc{R}_\infty}^{-1}$ and $e^{-\ovl{\log{\mc{R}_\infty}}}$ scale in the same way as a function of system size
($\ovl{\mc{R}_\infty}^{-1}\sim e^{-\ovl{\log{\mc{R}_\infty}}}$).
In the multifractal phase $\ovl{\mc{R}_\infty}^{-1}$ and $e^{-\ovl{\log{\mc{R}_\infty}}}$ scale in a different way,
$\ovl{\mc{R}_\infty}^{-1} \sim e^{-\alpha \ovl{\log{\mc{R}_\infty}}}$ with $\alpha = D_2/D_{\text{typ}} < 1$. This difference is also possible
to observe in the probability distribution of $\mc{R}_\infty$ (for the scaling with $L$ of the probability distribution of $\mc{R}_\infty$ see~Appendix), which in the multifractal region becomes long-tailed giving the
discrepancy in the scaling between mean and typical values. In the ergodic phase the probability distribution of $\mc{R}_\infty$ is
close to Gaussian. The inset of Fig.~\ref{fig:fig1_RP+PLBM}(c) shows $e^{\ovl{\log{IPR_x}}}$ as a function of $e^{\ovl{\log{\mc{R}_\infty}}}$ in a linear scale, giving indication that $e^{\ovl{\log{\mc{R}_\infty}}}\sim e^{\ovl{\log{IPR_x}}}$.
Furthermore, \rev{substituting the the equality
$\ovl{\mc{R}_\infty} = \ovl{IPR_x}$~(\ref{eq:relation}) in the latter one obtains} that $\ovl{IPR_x}^{-1} \sim e^{-\alpha \ovl{\log{IPR_x}}}$ with the same $\alpha$ as in $\ovl{\mc{R}_\infty}^{-1} \sim e^{-\alpha \ovl{\log{\mc{R}_\infty}}}$.

In the RPRM both $\ovl{\mc{R}}(t)$ and $e^{\ovl{\log{\mc{R}(t)}}}$ decay exponentially in time
in the non-ergodic phase, $1<\gamma <2$, $\ovl{\mc{R}}(t) \sim e^{\ovl{\log{\mc{R}(t)}}} \sim e^{-E_{Th} t}$
and polynomially with oscillatory time-dependence in ergodic phase, $\gamma<1$,
$\ovl{\mc{R}}(t) \sim e^{\ovl{\log{\mc{R}(t)}}} \sim [J_1(2\delta E t)/(2\delta E t)]^2$~\cite{deTomasi_2018}.
Here $J_1$ is the Bessel function of the first kind, $E_{Th}$ is the Thouless's energy and $2\delta E$ coincides in this case with the energy bandwidth $E_{BW}$ (as we take $\hat{P}_{\Delta E} = \mathds{1}$ for this model).
\rev{Some of authors of this paper have also studied in~\cite{deTomasi_2018} an accurate extraction of the $L$-dependence of $E_{Th}$ from $\mc{R}(t)$ nearly free from the finite size effects.}

Figure~\ref{fig:fig1_RP+PLBM}(b) shows $\ovl{\mc{R}}(t)$ and $e^{\ovl{\log{\mc{R}(t)}}}$ \rev{versus $t$} in the fractal critical region.
It gives an evidence that both $\ovl{\mc{R}}(t)$ and $e^{\ovl{\log{\mc{R}(t)}}}$ decay exponentially in time with the same rate $E_{Th}$.
The same dependence with time between mean and typical implies that their saturation values \rev{scale to zero as functions of $L$ in the same way.}
Figure~\ref{fig:fig1_RP+PLBM}(d) shows $\ovl{\mc{R}_\infty}^{-1}$ as a function of $e^{-\ovl{\log{\mc{R}_\infty}}}$
\rev{both in ergodic and
in fractal phases. In both phases typical and mean return probabilities scale in the same manner $\ovl{\mc{R}_\infty}^{-1} \sim e^{-\ovl{\log{\mc{R}_\infty}}}$, confirming above mentioned arguments about the fractal states.}

\begin{figure*}[t]
 \includegraphics[width=0.95\textwidth]{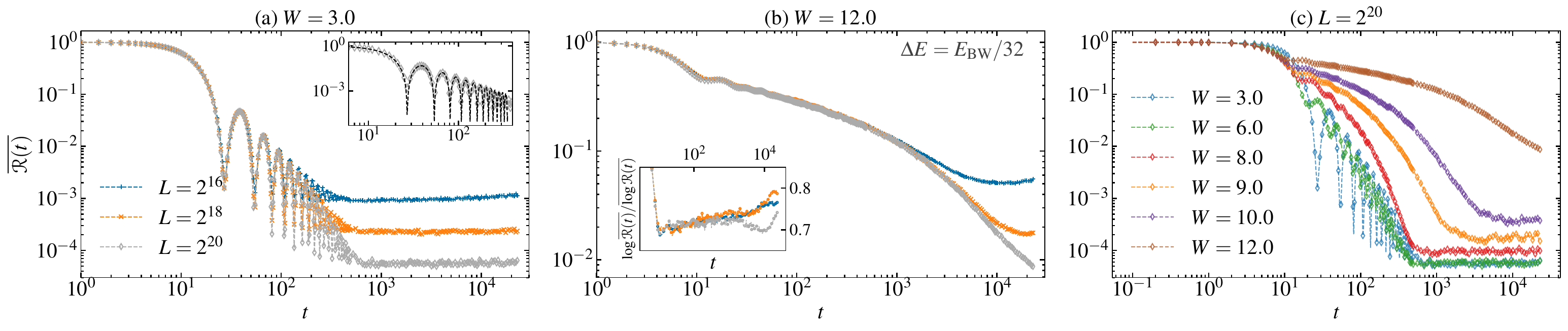}
 \caption{\rev{\bf The return probability $\ovl{\mc{R}}(t)$ versus $t$ for the RRG} with the energy shell $|\Delta E|=2\delta E = E_{BW}/32$ being a fraction of the whole energy bandwidth $E_{BW}$ for several disorder strengths $W$ and system sizes $L$.
 (a)~$\ovl{\mc{R}}(t)$ in the ergodic phase $W=3$ for several system sizes $L = 2^{16}, 2^{18}, 2^{20}$;
 The inset shows that $\ovl{\mc{R}}(t)$ for $W=3$ and $L=2^{20}$ almost coincides with the ergodic solution~\eqref{eq:R_erg} $\mc{R}_{erg}(t) = \left[\sin(2\delta E t)/(2\delta E t)\right]^2$ (shown by a black dashed line)~\cite{deTomasi_2018,tavora2017power,torres2018generic,santos2018nonequilibrium};
 (b)~$\ovl{\mc{R}}(t)$ in the non-ergodic phase $W=12$ at several system sizes $L = 2^{16}, 2^{18}, 2^{20}$.
 The inset shows a nontrivial plateau in the ratio $\log{\ovl{\mc{R}}}(t)/\ovl{\log{\mc{R}}}(t)<1$.
 (c)~$\ovl{\mc{R}}(t)$ for a fixed $L=2^{20}$ and several disorder strengths $W=3$, $6$, $8$, $9$, $10$, $12$.}
 \label{fig:Fig2_R(t)_erg_stretch-exp}
 \end{figure*}

\section{Anderson model on RRG}
Having shown that the difference in the behavior between the mean and the typical value of $\mc{R}(t)$ can be used to distinguish
ergodic and multifractal phases, we now study $\mc{R}(t)$ in the RRG.
In the RRG the existence of the multifractal phase is under \rev{active debate because of} two issues: First, the existence of a correlation length $L_{\text{cor}}$ which diverges \rev{as $W$ approaches} the Anderson
transition ($L_{\text{cor}} \sim e^{c/\sqrt{W_{AT}-W}}$)~\footnote{Recently, it has been proposed
~\cite{Kra18} a different expression for $L_{\text{cor}}$, $L_{\text{cor}}\sim e^{c/(W_{AT}-W)}$}. For finite systems of size $L$ smaller than $L_{\text{cor}}$ the wave-functions
could share properties both of localized and ergodic states and thus they could be mistakenly classified as multifractal.
Second, even for $L>L_{\text{cor}}$, finite size corrections
for the IPR might be quite strong and $\ovl{\sum_x |\phi_E(x)|^4} \sim \log(L)^\eta L^{-D_2}$, for some $\eta$ could \rev{affect the accuracy of the extraction of the critical exponent $D_2$}. Thus, the calculation of $D_2$ ($D_2=1$ for ergodic, $D_2<1$ for non-ergodic) is an extremely challenging numerical problem.

The Anderson model on the RRG has a mobility edge, thus in our study we consider only the energies in the middle of the spectrum, choosing
$|\Delta E|=2\delta E = E_{BW}/32$, ensuring that all the states $\{\phi_E\}_{E\in \Delta E}$ share the same properties for our choice of the disorder strength~$W$.
We perform the time evolution using full diagonalization for small systems sizes $L \le 2^{14}$, and using Chebyshev integration technique~\cite{Wei06} for larger $2^{15}\le L \le 2^{20}$.
The projector $P_{\Delta E}$ has been constructed using full diagonalization for $L \le 2^{14}$, and \rev{using a truncated
Chebyshev expansion~\cite{Bera17} for larger $2^{15}\le L \le 2^{20}$}.

\begin{figure}[t]
\includegraphics[width=0.5\textwidth]{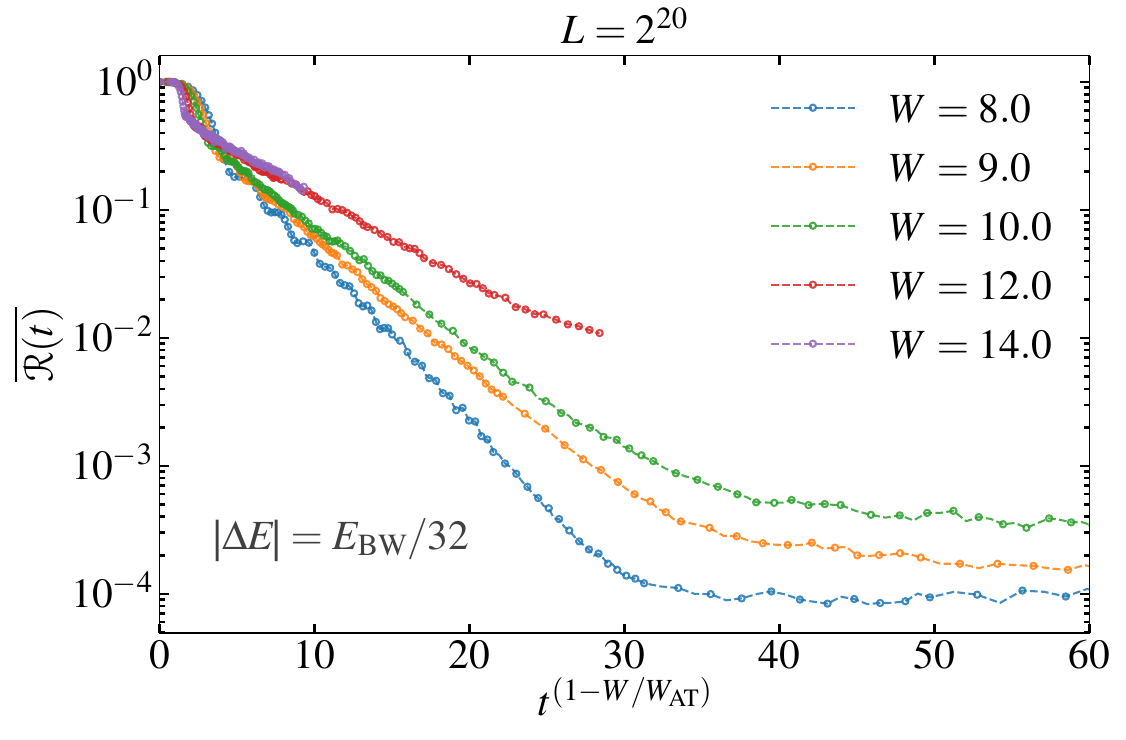}
\caption{
\rev{\bf The return probability $\overline{\mathcal{R}}(t)$ versus $t^{1-W/W_{AT}}$ in the non-ergodic phase} for several disorder strengths $W=8$, $9$, $10$, $12$ and fixed $L=2^{20}$ (rescaled data from Fig.~\ref{fig:Fig2_R(t)_erg_stretch-exp}(c)),
confirming the stretch-exponential behavior
$\overline{\mathcal{R}}(t)\sim e^{-\Gamma t^{\beta}}$, with $\beta=1-W/W_{AT}$ and $W_{AT}=18.2$.
}
\label{fig:Fig3_R(t)_vs_t_to_beta}
\end{figure}

Figure~\ref{fig:Fig2_R(t)_erg_stretch-exp}(a) demonstrates time dependence of the mean of $\mc{R}(t)$ \rev{at rather small} disorder strength
($W=3$) for several system sizes.
The presence of oscillations in the return probability $\mc{R}(t)$ surviving in the thermodynamic limit $L\to \infty$ confirms the existence of the \rev{fully ergodic phase consistent with Wigner-Dyson behavior}~\cite{deTomasi_2018,tavora2017power,torres2018generic,santos2018nonequilibrium}.
The inset of Fig.~\ref{fig:Fig2_R(t)_erg_stretch-exp}(a) confirms the form of oscillations without any fitting parameter
\begin{gather}\label{eq:R_erg}
\mc{R}_{erg}(t) = \left[\frac{\sin(2\delta E t)}{2\delta E t}\right]^2 \ ,
\end{gather}
which is valid for small energy shell $\delta E\ll E_{BW}$, approximating the local density of states with a box function \rev{and uncovering the rigidity of the spectrum}.
However, at \rev{moderate} disorder strength $W=12$ the time dependence of $\ovl{\mc{R}}(t)$ (shown in Fig.~\ref{fig:Fig2_R(t)_erg_stretch-exp}(b))
demonstrates \rev{absence of oscillations and} a clear bending in a log-log scale, \rev{excluding the power-law decay. This time dependence} is consistent with a stretched-exponential
decay $\ovl{\mc{R}}(t)\sim A e^{-\Gamma t^{\beta}}$ up to a possible sub-leading polynomial prefactor \rev{(see also Fig.~\ref{fig:Fig3_R(t)_vs_t_to_beta})}.
Also the typical value of $\mc{R}(t)$ decays like a stretched exponential, $e^{\ovl{\log{\mc{R}}}(t)}\sim A_{\text{typ}} e^{-\Gamma_{\text{typ}} t^{\beta}}$.
This stretched-exponential time behavior holds for moderate values of disorder strength $W$ in the extended phase ($0.4 \lesssim W/W_{AT}\lesssim 1$) both for $\ovl{\mc{R}}(t)$ and $e^{\ovl{\log{\mc{R}}}(t)}$ (for additional data for $\mc{R}(t)$ see~Appendix).

For disorder strengths between $ 0.16 \lesssim  W/W_{AT}\lesssim 0.4$  \rev{ergodic} oscillations are also present.
Nevertheless, their amplitudes reduce with the increasing system size,
preserving us from giving a final conclusion on the existence of the fully ergodic or multifractal phases in this regime.
Indeed, in~\cite{deTomasi_2018} it has been shown that in the fractal phase of the RPRM oscillations are just a finite\rev{-size} effect which disappears in the thermodynamic limit,
thus \rev{one cannot} exclude that \rev{the similar behavior could happen also for the AM on the RRG}.
Figure~\ref{fig:Fig2_R(t)_erg_stretch-exp}(c) shows $\ovl{\mc{R}}(t)$ for a fixed system size $L=2^{20}$ for several values of $W$.
The inset of Fig.~\ref{fig:Fig2_R(t)_erg_stretch-exp}(b) shows \rev{the ratio} $\log{\ovl{\mc{R}}}(t)/\ovl{\log{\mc{R}(t)}}$ for $W=12$, which develops a large plateau over more than 2 orders of magnitudes of $t$, increasing with increasing system size.
The formation of this plateau gives an evidence that the power $\beta$ is the same for mean and typical. Since value of the ratio $\log{\ovl{\mc{R}}}(t)/\ovl{\log{\mc{R}(t)}}$ at the plateau is less than unity one can claim that $\Gamma < \Gamma_{\text{typ}}$, as for large time $\log{\ovl{\mc{R}}}(t)/\ovl{\log{\mc{R}(t)}}\sim \Gamma/\Gamma_{\text{typ}}$.
This difference between $\Gamma$ and $\Gamma_{\text{typ}}$ has been observed for values $W>10$, while for smaller values of $W$ the numerics gives an evidence that $\Gamma =\Gamma_{\text{typ}}$~(see Appendix).

For disorder strengths between $0.4 \lesssim W/W_{AT}\lesssim 0.7$, where both the residual oscillations and the proximity to AT do not matter, the stretched-exponential parameter $\beta$ decays approximately linearly.
The linear extrapolation of $\beta(W)$ gives reasonable values of the Anderson localization transition $W_{AT}$, where $\beta(W_{AT})=0$.
Although at small disorder strength the ergodic oscillations of $\ovl{\mc{R}(t)}$ \eqref{eq:R_erg} hide the stretch-exponential behavior \rev{(as in RPRM model close to $\gamma=1$~\cite{deTomasi_2018}).
T}he linear extrapolation to this region $\beta(W\to 0)=1$ 
is consistent with works on classical diffusion on the Bethe lattice~\cite{Cassi1989random,Chinta2015heat}).

\rev{To avoid any problems with unstability of multi-parameter fit, in Fig.~\ref{fig:Fig3_R(t)_vs_t_to_beta} we show} $\ovl{\mc{R}}(t)$ exponentially decaying as a function of $t^{(1-W/W_{AT})}$ for several values of $W$ providing \rev{the direct} indication that $\beta~\simeq (1-W/W_{AT})$.
In summary, our analysis for $\ovl{\mc{R}}(t)$ provides the evidence that for small disorder $W<0.16 W_{AT}\simeq 3$ the RRG is in the \rev{fully} ergodic phase (due to the oscillatory behavior \eqref{eq:R_erg}), while at $0.4<W/W_{AT}<0.7$ the behavior is certainly \rev{non-ergodic} (absence of oscillations, $\Gamma<\Gamma_{typ}$, stretch-exponential time-dependence of $\ovl{\mc{R}}(t)$).
As a result, one should expect the ergodic transition in the range $0.16<W_{EMT}/W_{AT}<0.4$ in agreement with \cite{Alt16}.

\begin{table*}[t]
 \begin{tabular}{ l | c | c | c | c }
 & Ergodic
 \footnote{
 PLRBM $(a<1)$, RPRM $(\gamma<1)$, and RRG $W\lesssim 3$.
 $a_0=2(3)$ for $|\Delta E|\ll E_{BW} (|\Delta E|\simeq E_{BW})$.
 Oscillations of the type \eqref{eq:R_erg} are present in both cases.}
 &PLRBM $(a=1)$
 & RPRM ($1<\gamma<2$)
 & RRG ($8\lesssim W\lesssim12$)\\
 \hline			
 $\ovl{\mc{R}(t)}$ & $\sim t^{-\alpha_0}$ & $\sim t^{-\alpha_1}$ & $\sim e^{-E_{Th} t}$ & $\sim e^{-\Gamma t^\beta}$ \\
 $\exp[\ovl{\log \mc{R}(t)}]$ & $\sim t^{-\alpha_0}$ & $\sim t^{-\alpha_2}$ & $\sim e^{-E_{Th} t}$ & $\sim e^{-\Gamma_{\text{\text{typ}}} t^\beta}$ \\
 $\log(\ovl{\mc{R}(t)})/\ovl{\log \mc{R}(t)}$ & 1 & $\alpha_1/\alpha_2<1$ & 1 & $\Gamma/\Gamma_{\text{\text{typ}}}\leq 1$ \\
 \hline
 \end{tabular}
 \caption{Behavior of typical and average return probability and ratio of their logarithms for large $t$ for the different models.}
\end{table*}

It is important to underline that the time scale $t^* \sim \Gamma^{-1/\beta}$ in which the decay of $\ovl{\mc{R}}(t)$ can be distinguished from an algebraic decay diverges approaching the Anderson transition \rev{(see also~\cite{Tikhonov_misc})} i.e. for $W=14$ the bending in a log-log plot is only visible for $t^\star\approx 10^{4}$ and it requires having system size of $L=2^{20}$, thus the decay of $\ovl{\mc{R}}(t)$ for smaller times and smaller system size could be
interpreted as a power law~\cite{Biro17}.
Moreover, in a recent work~\cite{Biro17} it is argued on the base of numerics, that a possible power law decay of
$\ovl{\mc{R}}(t)\sim t^{-\zeta}$ is consistent with an algebraic dependence of the overlap of different wavefunctions $\mc{K}(\omega)\sim \omega^{1-\zeta}$ defined as
\begin{equation}
 \mc{K}(\omega) = \frac{1}{\mc{N}}\ovl{\sum_{E,E'\in \Delta E} |\langle x| E\rangle |^2 |\langle E'| x\rangle |^2 \delta (\omega - E+E')}
\end{equation}
with a normalization constant $\mc{N}$ ensuring $\int d\omega \mc{K}(\omega) = 1$.

However, using stationary phase approximation it is possible to show that for $\ovl{\mc{R}}(t)\sim e^{-\Gamma t^{\beta}}$, the overlap decays as $\mc{K}(\omega) \sim \omega^{-\frac{1+(1-\beta)^{-1}}{2}}$ for moderately large $\omega$ and as $\mc{K}(\omega) \sim \omega^{-(1+\beta)}$ for very large $\omega$.
As for observed values of $\beta\lesssim 0.5$ the difference between above mentioned exponents is less than $7~\%$, a stretched-exponential behavior for $\ovl{\mc{R}}(t)$ can be consistent, in \rev{the} first approximation, with a single power law behavior of $\mc{K}(\omega)$ observed in other works (e.g.,~\cite{Biro17,deLuca_misc}) \rev{as well as with the power-law with logarithmic corrections~\footnote{In the work~\cite{Tikhonov_misc} authors claim that at large $\omega$ the overlap correlation function decays as $\mc{K}(\omega)\sim \left[\omega \log^{3/2}(1/\omega)\right]^{-1}$}.}

\section{Classical random walk approximation}
Let us now present a classical model of subdiffusion which can explain stretched-exponential behavior of the return probability on the RRG, while giving normal sub-diffusion on a regular lattice. Let us consider a random walk in which every time $\Delta t$ the walker makes a step in a randomly picked direction.
\rev{
We assume that different dwelling times $\Delta t$ and jump directions, determining the random number $N(t)$ of steps the walker takes in time $t$, are statistically independent. Within this assumption on a \emph{line}, one can easily see that the averaged square distance from the initial point
is determined solely by $N(t)$}
\begin{equation}
\langle x^2(t)\rangle=a^2 N(t),
\label{eq:xsubline}
\end{equation}
where $a$ is the lattice constant. If \rev{we have} $N(t)\sim (t/\tau)^\beta$ we \rev{straightforwardly obtain} a subdiffusion law
\begin{equation}
\langle x^2(t)\rangle=a^2 (t/\tau)^\beta.
\label{eq:xsubline}
\end{equation}
It is possible to see that, by choosing $P(\Delta t)\sim 1/\Delta t^{1+\beta}$ for $\beta\leq 1$,
the typical number of steps in an interval is \rev{indeed scaling as} $N(t)\sim t^\beta$. For $\beta>1$, instead we have $N\sim t$~\footnote{It did not escape to us the relation between this phenomenology and the ``miniband'' phenomenology of~\cite{Kra18}. Unfortunately we are not able to provide a quantitative connection between the two and we plan to investigate this point in the future.}.

On any regular lattice, the probability distribution follows the Markov rate equation
\begin{equation}
P(y,N+1)-P(y,N)=\sum_{x}(\mathbb{A}-(K+1)\mathbb{I})_{y,x}P(x,N),
\label{eq:MarkP}
\end{equation}
where $K+1$ is the connectivity and $\mathbb{A}$ the adjacency matrix of the graph. While for a typical non-expander like a square lattice $\mathbb{Z}^d$, or similar, the decay of the return probability after $N$ steps is power-law $\mc{R}(N)\sim N^{-d/2}$, for a RRG/Bethe lattice the return probability scales exponentially
\begin{equation}
\mc{R}(N)=P(x,N|x,0)\sim e^{-g N},
\label{eq:}
\end{equation}
irrespective of $x$, where $g$ is the gap in the adjacency matrix, $g=(K+1)-2\sqrt{K}$.
If we now \rev{take into account the above mentioned assumption} $N(t)\sim (t/\tau)^\beta$ we \rev{immediately obtain} a stretched-exponential form
\begin{equation}
\mc{R}(t)\sim e^{-\Gamma t^\beta},
\label{eq:}
\end{equation}
which is in accord with our numerics.
\rev{
Moreover, for a RRG/Bethe lattice
the averaged distance from the initial point
grows linearly with $N(t)$
\begin{equation}
\langle x(t)\rangle=a N(t),
\label{eq:x_RRG_N(t)}
\end{equation}
giving in our case subballistic wavepacket spreading
\begin{equation}
\langle x(t)\rangle=a (t/\tau)^\beta \ .
\label{eq:x_RRG_t}
\end{equation}

The prediction of the subdiffusive spreading of the wavepacket following from \eqref{eq:xsubline} needs to be verified in further works.}
Notice moreover, that the fluctuations of $N(t)$ can explain the difference between $\log\ovl{\mc{R}}(t)$ and $\ovl{\log(\mc{R}(t))}$, even if the distribution of $N(t)$ does not have long tails. For example a distribution like $P(N)=\frac{e}{\nu}e^{-N/\nu}$ for $N\geq \nu$ and 0 otherwise gives a ratio $\log\ovl{\mc{R}}(t)/\ovl{\log(\mc{R}(t))}=1/2$.
\rev{The above mentioned assumption of the classical dynamics in RRG can be justified in our numerics due to not too small width $\Delta x\sim \hbar v/|\Delta E|$ of the initial wavepacket $\hat{P}_{\Delta E}|x\rangle$.
The power-law distribution of the dwelling times $\Delta t$ is possibly related to the strength of the on-site disorder.
}

 \begin{figure*}[t]
 \includegraphics[width=0.95\textwidth]{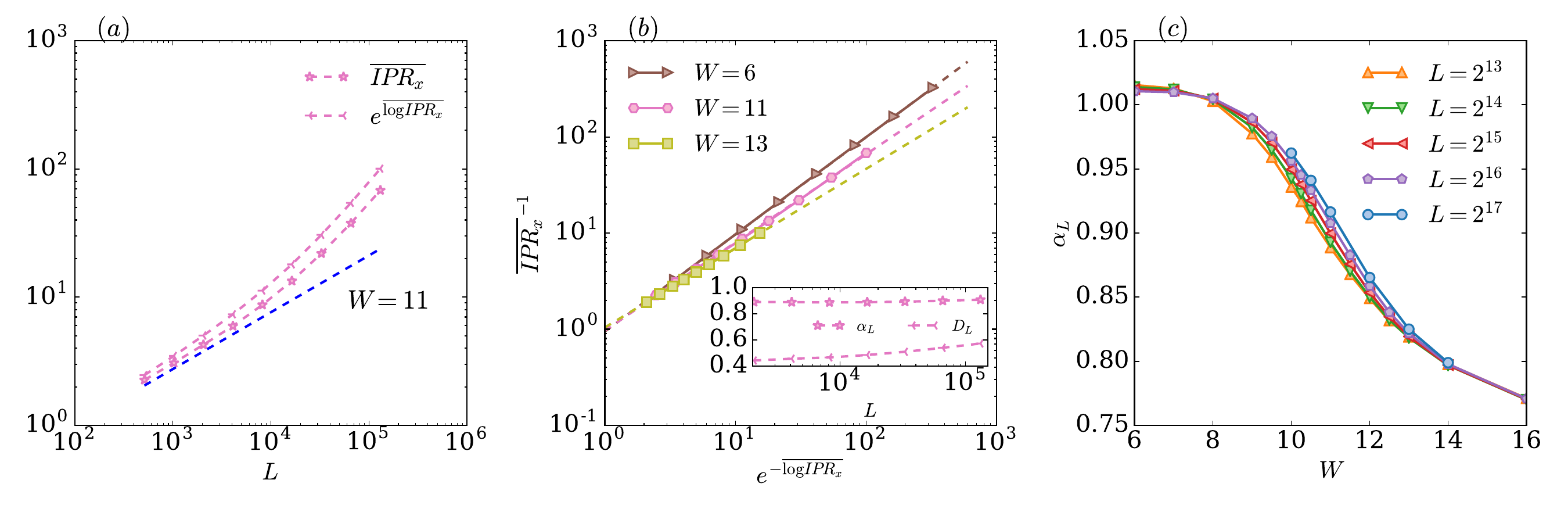}
 \caption{\rev{\bf Scaling of mean and typical IPR in RRG versus the system size $L$ and versus each other.}
 (a)~$\ovl{IPR_x}$ and $e^{\ovl{\log{IPR_x}}}$ as a function of $L$ \rev{at the} disorder strength $W=11$.
 \rev{Both curves show significant deviations from the power-law scaling $IPR_x \sim L^{-D}$.}
 (b)~$\ovl{IPR_x}^{-1}$ \rev{versus} $e^{-\ovl{\log{IPR_x}}}$ ($\ovl{IPR_x} \sim e^{\alpha \ovl{\log{IPR_x}}}$) for several $W$.
 \rev{Different points of the curve correspond to the} different system sizes $L=2^9-2^{17}$.
 Panel~(c) shows \rev{$L$-dependent power $\alpha_L$ of the mutual IPR scaling} as a function of $W$. $\alpha_L$ has been extracted from the linear fitting of $\log{\ovl{IPR_x}}$ versus $\ovl{\log{IPR_x}}$ with an enlarging $L$. The inset of panel~(b) shows $\alpha_L$ and $D_L$ as a function of $L$. $D_L$ has been extracted from the linear fitting of $\left(-\log{\ovl{IPR_x}}\right)$ versus $\log L$.
 \rev{The subscript $L$ in $\alpha_L$ and $D_L$} indicates the largest system size considered in the fit starting with $L=2^9$.
 }
 \label{fig:Fig4_IPR_typ_mean}
 \end{figure*}

\section{Participation ratios}
\rev{In this section we analyze the system size dependence of the saturation values \eqref{eq:relation} of mean $\ovl{\mc{R}_\infty}=\ovl{IPR_x}$ and typical $e^{\ovl{\log{\mc{R}_\infty}}}\sim e^{\ovl{\log{IPR_x}}}$ return probability in AM on RRG.}
We \rev{directly observe scaling of IPRs with system size $IPR_x \sim L^{-D_2}$, but the scaling exponents $D_2$} have not yet reached saturation.

Using the time evolution algorithm it is difficult to extract the saturation values of $\mc{R}(t)$ systematically and reliably, since very large times are needed. Therefore, we find it easier to analyze $IPR_x$ using a shift-inverse exact diagonalization technique.
Figure~\ref{fig:Fig4_IPR_typ_mean}(a) show $\ovl{IPR_x}$ and $e^{\ovl{\log{IPR_x}}}$
as a function of $L$ in a log-log scale for a fixed disorder strength $W=11$.
Strong finite size effects are visible for available
systems sizes, which makes the extrapolation of $D_2$ and $D_{\text{\text{typ}}}$ \rev{unreliable.} 
Nevertheless, $\ovl{IPR_x}$ and $e^{\ovl{\log{IPR_x}}}$ seem to suffer from similar finite-size effects.
Indeed, plotting $\ovl{IPR_x}$ parametrically as a function of $e^{\ovl{\log{IPR_x}}}$ drastically
reduces finite-size effects.
Figure~\ref{fig:Fig4_IPR_typ_mean}(b) shows $\ovl{IPR_x}$ as a function of $e^{\ovl{\log{IPR_x}}}$ for several values of $W$, giving \rev{an} indication that $\ovl{IPR_x}\sim e^{\alpha\ovl{\log{IPR_x}}}$.
As we have already shown, in the ergodic phase $\alpha = 1$, while in the multifractal phase one expects $\alpha=D_2/D_{\text{\text{typ}}}<1$.
Using an enlarging linear fitting procedure we are able to extract the exponent $\alpha$ as a function of \rev{the disorder strength and the} system size $L$, $\alpha_L$, (\rev{here $L$ indicates} the last system size \rev{which} has been taken in consideration in the fit starting from $L=2^9$).
\rev{Extracting in the same manner the critical exponent $D$ of the $L$-scaling of the mean $IPR_x$ value, we compare the results for $\alpha_L$ and $D_L$ in the} inset of Fig.~\ref{fig:Fig4_IPR_typ_mean}(b). 
\rev{From the inset one can see that} $D_L$ has a change of $30~\%$ for available system sizes, while $\alpha_L$ changes only by $3~\%$.
\rev{Nevertheless, $\alpha_L$ increases with $L$} and we cannot in principle exclude that its asymptotic value may be $\alpha=1$.

Figure~\ref{fig:Fig4_IPR_typ_mean}(c) shows $\alpha_L$ for several $L$ as a function of $W$. For $W<10$, $\alpha\approx 1$ \rev{gives} the evidence that in this regime even at our system sizes, the eigenstates are ergodic \rev{(but possibly not fully ergodic as at $W\lesssim 3$)}. For $W>10\approx W_{EMT}$, $\alpha$ drops to a smaller value confirming that for available systems sizes the system has not developed ergodicity. The flow of
$\alpha_L$ towards unity with increasing $L$ is visible at least for $10<W<12$, while for $W>12$ the data does not change with system size.
For large disorder strength one should also keep in mind that the convergence could be \rev{induced by the finite system sizes, which are} small compared to the correlation length $L_{\text{cor}}$
\footnote{The difference between mean and typical $\mc{R}(t)$ is also possible to observe in the probability distribution of $\mc{R}_\infty$. For the scaling with $L$ of the probability distribution of $\mc{R}_\infty$ see~Appendix. }.

\section{Conclusion}
We have studied the quantum dynamics of 
particle \rev{initially prepared in a narrow wave-packet form}
in three different ensembles of disordered systems, giving a characterization
of multifractal phases based on the statistics of the return probability. In particular, we have studied the return probability $\mc{R}(t)$ to the initial \rev{state} during the quantum dynamics. We have proposed that in the multifractal phase, fluctuations over disorder and initial site are so strong that the long time limit of the mean and typical value of $\mc{R}(t)$ scale to zero differently as a function of system size. In the ergodic \rev{and in fractal phases} the scaling is \rev{shown to be} the same.

First, we have benchmarked these ideas in the power law random banded matrix ensemble, where one observes both
\rev{ergodic and multifractal phases}. We have shown that the long time limit of the mean and typical value of $\mc{R}(t)$ scale to zero in the same way in its ergodic regime, while at criticality, where all the states are multifractal, the scaling of these observable\rev{s} differ from each other.

Second, we have pointed out, analyzing the Rosenzweig-Porter random matrix model, that this difference in the scaling disappears in the case of fractal \rev{(but not multifractal)} states.

Finally, we have used this idea to tackle the Anderson model on random regular graph, in which the existence of an extended multifractal phase is under debate.
We present \rev{the results for the return probability} $\mc{R}(t)$ for system sizes\rev{, where the convergence of $\mc{R}(t)$ is ensured},
and provide \rev{the} numerical evidence of the difference of the mean and typical values of the return probability, \rev{giving} a signature of \rev{non-ergodic behavior} of eigenstates, in the range $0.4<W/W_{AT}<0.7$.
Furthermore, we have shown that in this range $\mc{R}(t)$ decays
likes a stretched exponential
and we have extracted the parameters of this stretched-exponential decay.
We give a phenomenological classical subdiffusive hopping model which reproduces the stretched exponential of the return probability
\rev{and provides predictions of the wavepacket evolution with time, which are worth to further verify}.

For small disorder strengths, $W\lesssim 0.16 W_{AT}$, $\mathcal{R}(t)$ shows oscillations which survive in the thermodynamic limit, confirming the existence of the \rev{fully} ergodic phase \rev{consistent with the standard Wigner-Dyson behavior}.

Our analysis based on dynamical properties allow us to
conclude that the RRG is in the \rev{fully-}ergodic phase at least for small disorder $W<0.16 W_{AT}$
and in non-ergodic extended phase at least for the $0.4<W/W_{AT}<0.7$.
What implies that a transition between \rev{ergodic and non-ergodic phases} should exist in the range $0.16<W_{EMT}/W_{AT}<0.4$ in agreement with \cite{Alt16}.

The subdiffusion results for the wavepacket spreading \eqref{eq:xsubline} guessed from the classical model need more analysis and we left them for further investigations.

{\bf Acknowledgements.}
The authors would like to thank \rev{A.~L.~Burin, F.~Evers,} M.~Heyl, V.~E.~Kravtsov, G.~Lemari\'e, F.~Pollmann and \rev{K.~S.~Tikhonov} for many interesting discussions.
A.S.\ is partially supported by a Google Faculty Award.
S.B. acknowledges support from DST, India, through Ramanujan Fellowship Grant No. SB/S2/RJN-128/2016.
I.M.K. acknowledges the support of
German Research Foundation (DFG) Grant No. KH~425/1-1 and the Russian Foundation for Basic Research.

\rev{{\it Note added.} During the consideration of the paper in the journal, authors have become aware of the work~\cite{Tikhonov_misc} 
considering, in particular, the return probability $\mc{R}(t)$ in RRG and having similar results of the stretched-exponential decay of $\mc{R}(t)$
with time in the corresponding range of system sizes and time intervals.}

\bibliography{RRG_bib}
\begin{widetext}
\appendix

\section{Power-law banded random matrix}
In this section we provide additional data for the power-law banded random matrix~(PLBM).

\subsection{Return Probability}

\begin{figure}[htb]
  \centering
 \includegraphics[width=0.7\columnwidth]{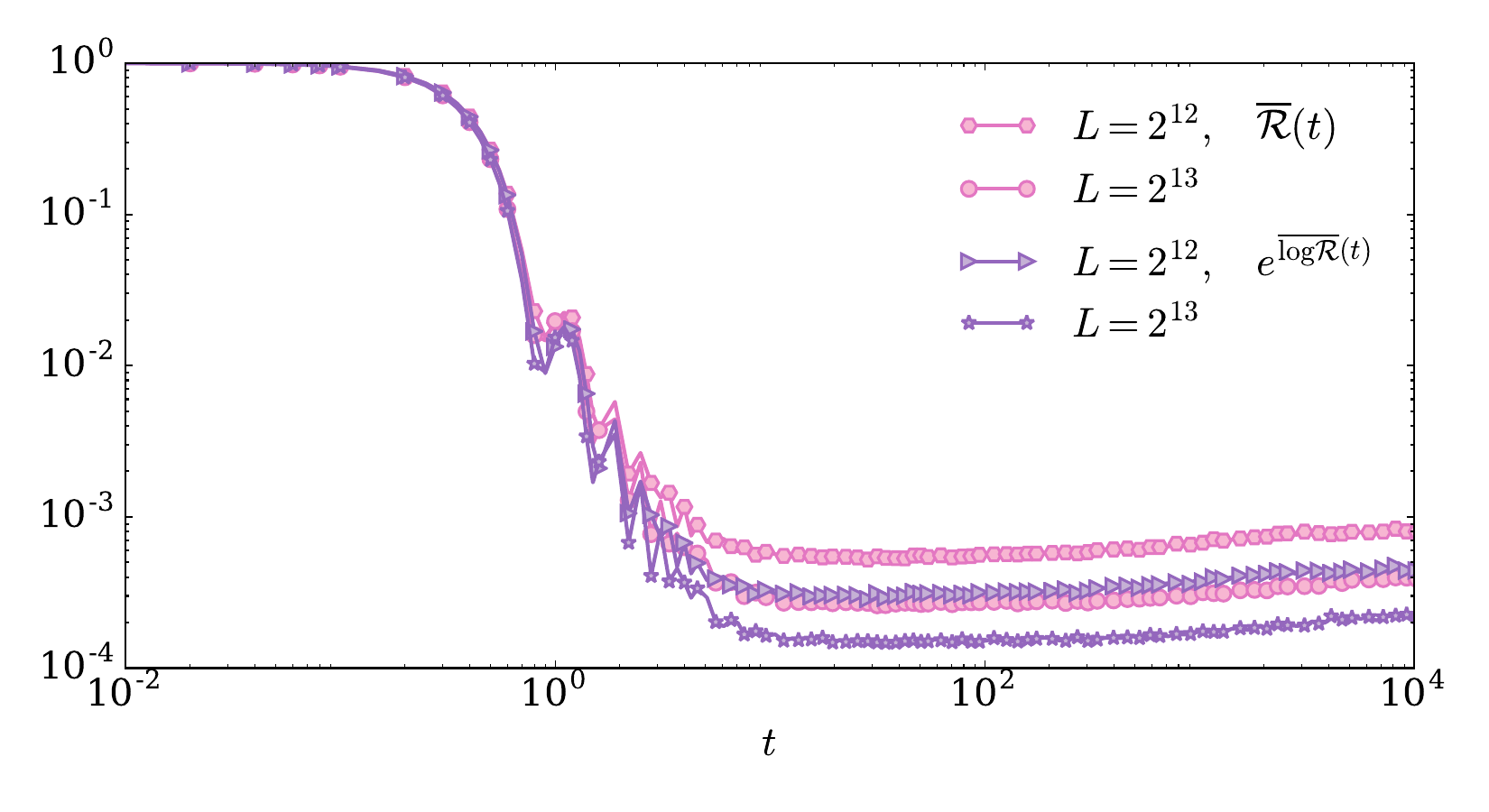}
 \caption[Return probability for the power-law banded matrix in the ergodic phase]{The figure shows $\ovl{\mc{R}(t)}$ and
$e^{\ovl{\log{\mc{R}}(t)}}$
 in the ergodic phase ($a=0.5$, $b=1$) of the PLBM for two systems sizes. See the main text for additional details of the model parameters.}
 \label{fig:PLBM_time}
 \end{figure}

In this section we study the decay of $\mc{R}(t)$ with time for the PLBM.
We perform the time evolution using exact diagonalization method as mentioned in the main text.
The decay of the return probability at the critical point is already shown in the main
text, here we show a similar data in the ergodic phase of the model.
At the critical point of the PLBM, where all states are multifractal, both $\ovl{\mc{R}}(t)$
and $e^{\ovl{\log{\mc{R}(t)}}}$ decay algebraically, but with different power laws.
%
%
Instead, in the ergodic phase ($a<1$)
$\ovl{\mc{R}}(t)$ and $e^{\ovl{\log{\mc{R}(t)}}}$ decay asymptotically at the same rate, as shown in Fig.~\ref{fig:PLBM_time}.
As a consequence of a different rate of decay between $\ovl{\mc{R}}(t)$ and $e^{\ovl{\log{\mc{R}(t)}}}$, in the multifractal phase
the saturation values $\ovl{\mc{R}_{\infty}}$ and $e^{\ovl{\log{\mc{R}_{\infty}}}}$
can have different scaling to zero as a function of $L$, $\ovl{\mc{R}_{\infty}} \sim L^{-D_2}$
and $e^{\ovl{\log{\mc{R}_{\infty}}}} \sim L^{-D_{\text{typ}}}$ ($D_2<D_{\text{typ}}<1$),
while in an ergodic phase $D_2 = D_{\text{typ}} = 1$.
\begin{figure}[htb]
  \centering
 \includegraphics[width=0.7\columnwidth]{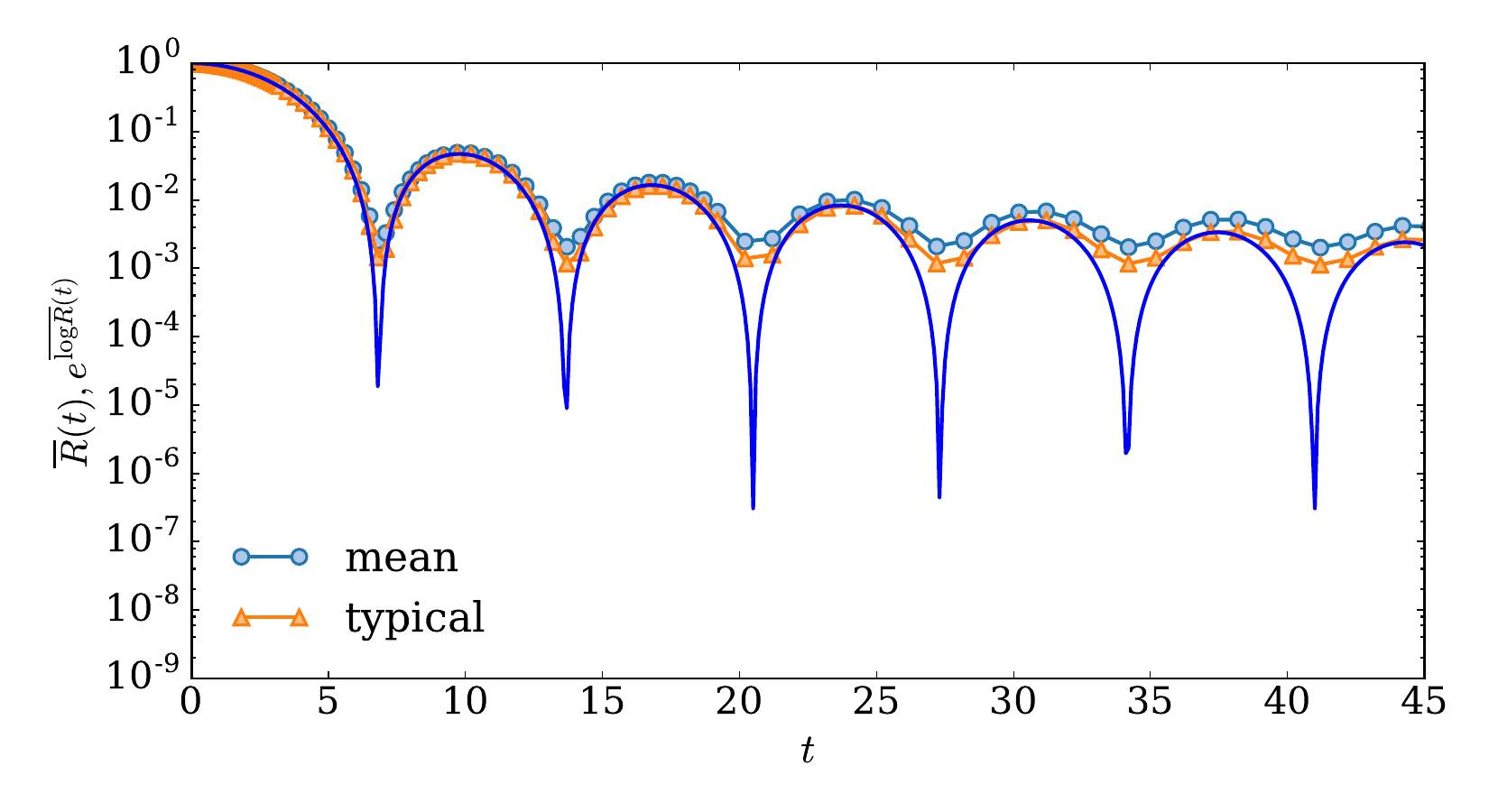}
 \caption[Return probability for the power-law banded matrix in the ergodic phase]{The figure shows $\ovl{\mc{R}(t)}$ and
$e^{\ovl{\log{\mc{R}}(t)}}$
 in the ergodic phase ($a=0.5$, $b=1$) of the PLBM for $L=2^{13}$, in this case we restricted the dynamics to energies in $\Delta E=[-\delta E, \delta E]$ with $|\Delta E|=2\delta E = E_{BW}/32.$ The solid line shows that
 $\mathcal{R} \sim \left[\sin(2\delta E t)/2\delta E t\right]^2$. }
 \label{fig:PLBM_time1}
 \end{figure}
 \begin{figure}[htb]
  \centering
 \includegraphics[width=0.7\columnwidth]{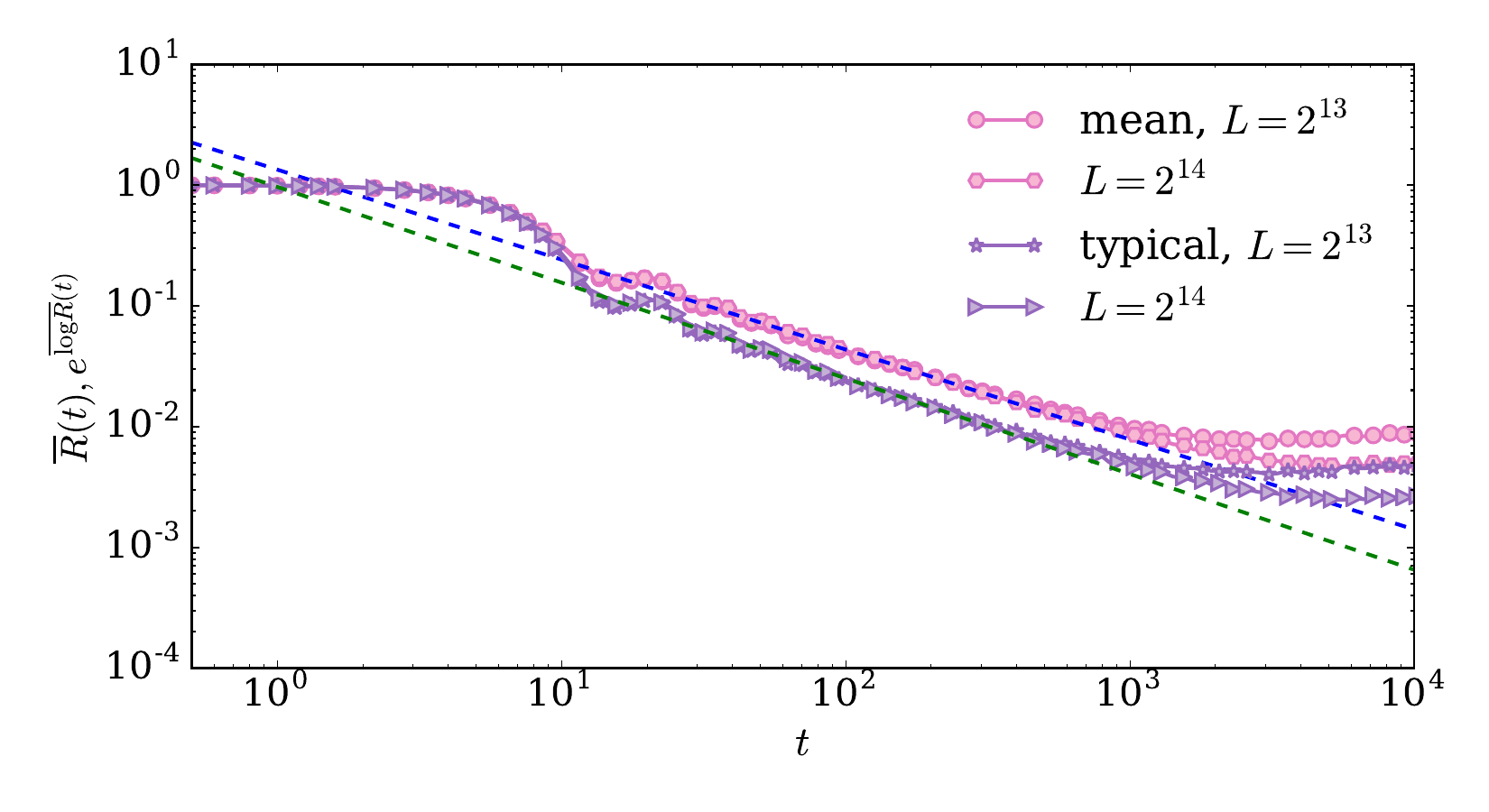}
 \caption{The figure shows $\ovl{\mc{R}(t)}$ and
$e^{\ovl{\log{\mc{R}}(t)}}$
 in the multifractal phase ($a=1$, $b=1$) of the PLBM for $L=2^{13}$, in this case we restricted the dynamics to energies in $\Delta E=[-\delta E, \delta E]$ with $|\Delta E| = 2\delta E = E_{BW}/32.$. Also in this case both the mean and the typical value decays algebraically in time
 but with different exponent.}
 \label{fig:PLBM_time2}
 \end{figure}

Figure~\ref{fig:PLBM_time1} shows $\ovl{\mc{R}}(t)$ and $e^{\ovl{\log{\mc{R}(t)}}}$ in the ergodic phase for the PLBM for one system size $L=2^{13}$,
in this case we projected onto eigenstates with energies within $\Delta E = [-\delta E , \delta E ]$ with $|\Delta E| = 2\delta E = E_{BW}/32$, where $E_{BW}$ is the energy band-width.
For these values ($a=0.5$, $b=1$) the system being ergodic provides the rigidity to the spectrum and its local density of states can be well approximated as a box function leading to $\mc{R}(t) \sim \left[\sin(2\delta E t)/2\delta E t\right]^2$. Moreover, as expected the mean and the typical value
of $\mathcal{R}(t)$ have
the same time dependence.
Figure~\ref{fig:PLBM_time2} shows $\ovl{\mc{R}}(t)$ and $e^{\ovl{\log{\mc{R}(t)}}}$ in the multifractal phase ($a=1$, $b=1$) for the PLBM for two system sizes $L=2^{13}, 2^{14}$, also here we projected on the energy shell $\Delta E=[-\delta E , \delta E]$ with $|\Delta E|=2\delta E = E_{BW}/32$.
In this phase there are no persistent oscillations with time and $\ovl{\mc{R}}(t)$ and $e^{\ovl{\log{\mc{R}(t)}}}$ decay with two different exponents ($\ovl{\mc{R}}(t) \sim t^{-\alpha_1}, e^{\ovl{\log{\mc{R}(t)}}} \sim t^{-\alpha_2})$ with ($\alpha_2>\alpha_1$),
indicating its multifractal properties.


\subsection{Probability distribution of IPR}
  \begin{figure}[htb]
  \centering
 \includegraphics[width=0.8\columnwidth]{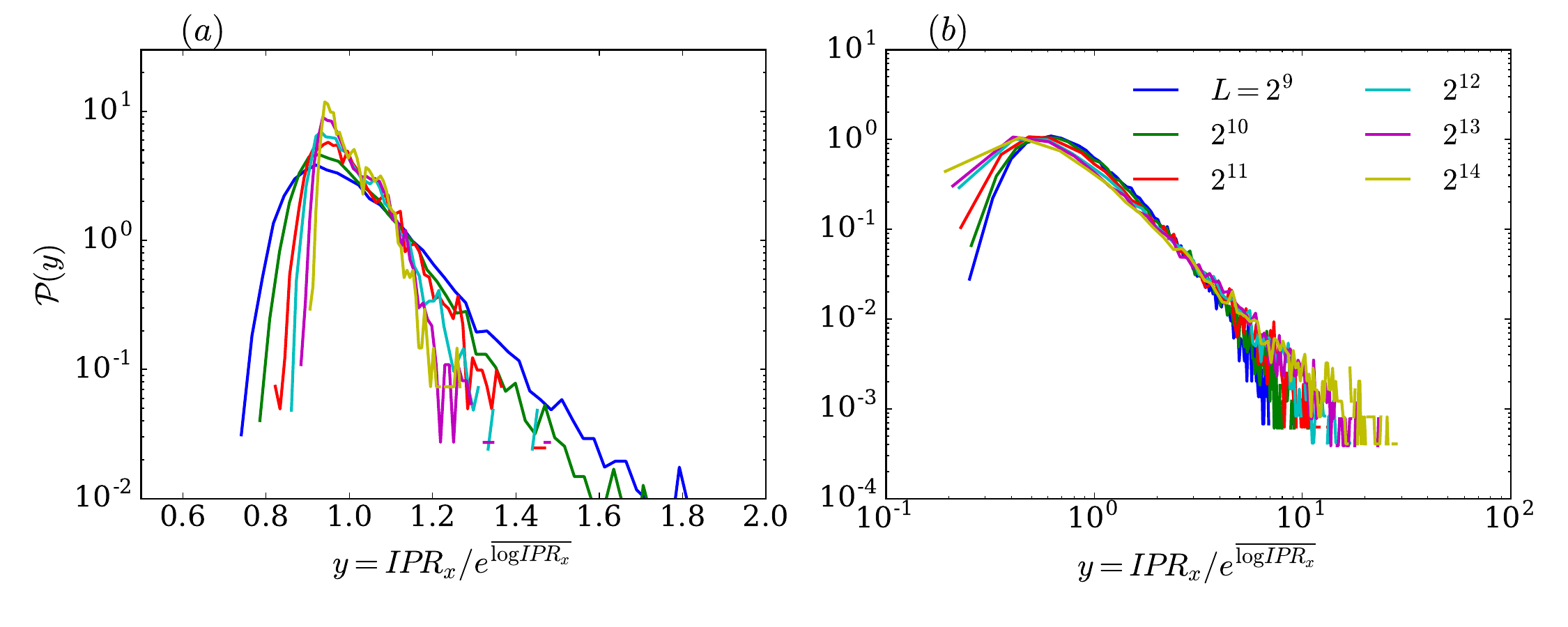}
 \caption[Probability distribution for the inverse participation ratio for the random regular graph] {Panel (a) shows the probability distribution
 $\mc{P}(y)$ for the rescaled random variable $y= \text{IPR}_x/e^{\ovl{\log{\text{IPR}_x}}}$
  for PLBM in the ergodic phase ($a=0.5$, $b=0.5$) for several system sizes. Panel (b) shows the probability distribution $\mc{P}(y)$ in the
 multifractal phase ($a=1$, $b=0.5$), in this case $\mc{P}(y)$ has power-law tails, indicating that $\ovl{\text{IPR}_x}^{-1} \sim
e^{-\alpha \ovl{\log{\text{IPR}_x}}}$ with $\alpha <1$.}
 \label{fig:Prob_PLBM}
 \end{figure}

The difference in scaling between the mean and typical value
originates from the tails of
the probability distribution of  the long time limit of the return probability, $\mc{R}(t \rightarrow \infty)$, which is the inverse
participation ratio~($\text{IPR}_x$).
In the multifractal region the probability distribution of  $\text{IPR}_x$ becomes long-tailed giving the
discrepancy in the scaling between mean and typical values as shown in Fig.~\ref{fig:Prob_PLBM}(b) for
the rescaled random variable $y= \text{IPR}_x/e^{\ovl{\log{\text{IPR}_x}}}$. This observation is consistent with our data in the main text
about the time dependence of the return probability.

While in the ergodic phase, Fig.~\ref{fig:Prob_PLBM}(a) the probability distribution of
$y$
has exponentially decaying tails and it
shrinks with increasing system size with a well defined mean close to 1. This validates again our previous observation in Fig.~\ref{fig:PLBM_time},
which is that at log time the two exponents $D_2 = D_{\text{typ}} = 1$, i.e.,   $\ovl{\text{IPR}_x}\sim
e^{\ovl{\log{\text{IPR}_x}}}$.

\section{Random Regular Graph}
In this section we provide additional data of return probability and the probability distribution of inverse participation ratio for Random Regular
Graph~(RRG).
\subsection{Return probability}
\begin{figure}[htb]
\centering
\includegraphics[width=0.9\columnwidth]{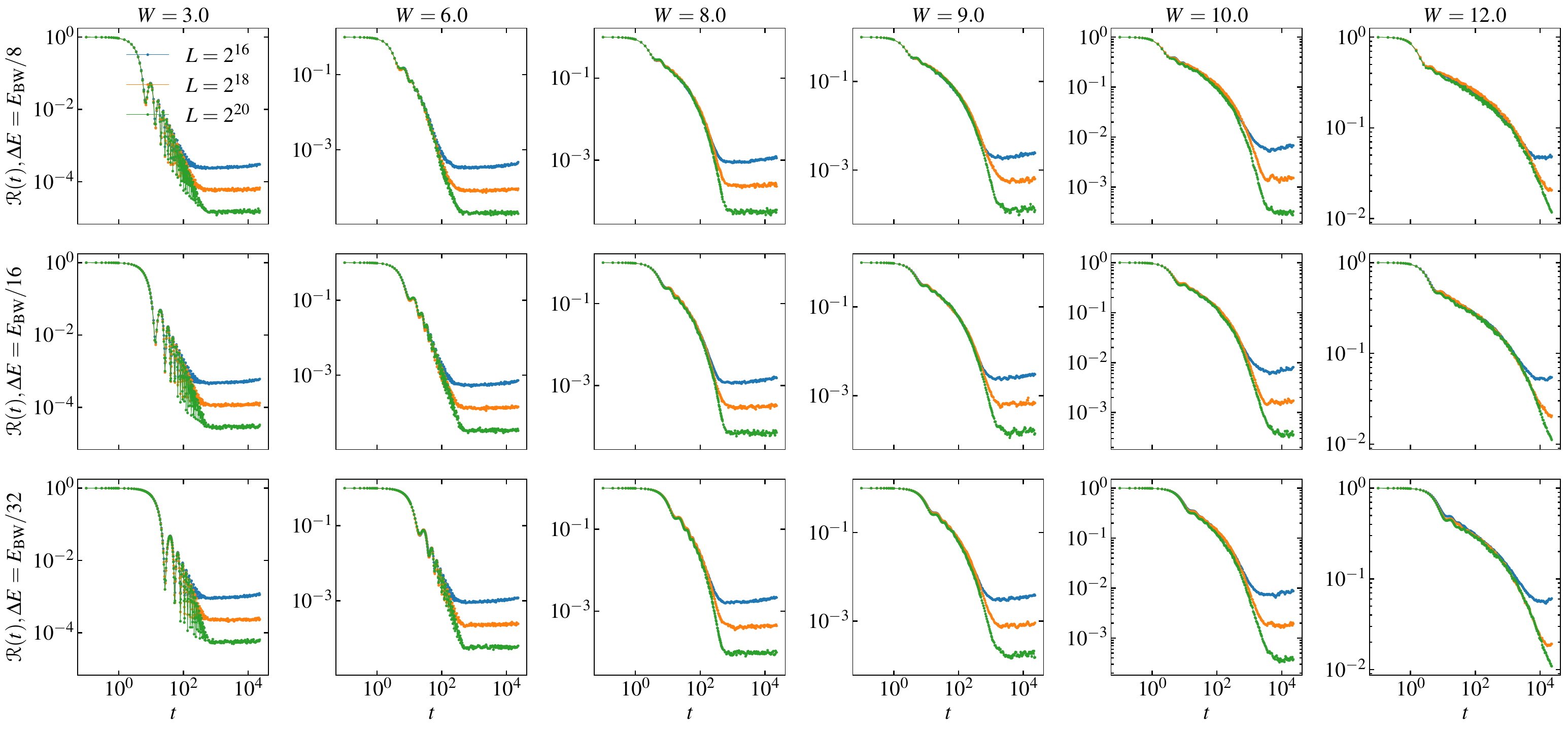}
\caption{The mean return probability $\ovl{\mc{R}}(t)$ for several $W$ and for the three largest system sizes $L=2^{16}, 2^{18}, 2^{20}$ and for several energies shells ($|\Delta E|/E_{BW} = 1/8$, $1/16$, $1/32$).}
\label{fig:Mean_rrg}
\end{figure}
Figure~\ref{fig:Mean_rrg} (Figure~\ref{fig:Typical_rrg}) shows $\ovl{\mc{R}}(t)$ ($e^{\ovl{\log{\mc{R}}}(t)}$) for several values of $W$ and for the three largest system sizes ($L= 2^{16}, 2^{18}, 2^{20}$) and for several $\Delta E$.
For $W\le 3$ and for all $|\Delta E| \le E_{BW}/8$ oscillations are present and the decay of $\overline{\mathcal{R}}(t)$ is consistent with $\left[\sin(2\delta E t)/2\delta E t \right]^2$, giving the evidence of the existence of an ergodic phase.
For $W=6$ and $W=8$ residual oscillations are also present and they are more pronounced for smaller $\Delta E$, but the decay of $\overline{\mathcal{R}}(t)$ is not consistent with a power-law ($\sim t^{-2}$) indicating that the phase could be not completely
ergodic. Nevertheless, it is important to point out that decreasing $\Delta E$ for this values of $W$ the curves show a behavior which seems to resemble a power-law decay (for $|\Delta E| = E_{BW}/32$), leaving open the possibility that the system is
ergodic but only in a smaller energy shell.
For larger values $W\ge 9$ the oscillations disappear completely and the decay of $\overline{\mathcal{R}}(t)$ is consistent with a stretched-exponential decay.
\begin{figure}[htb]
\centering
\includegraphics[width=0.9\columnwidth]{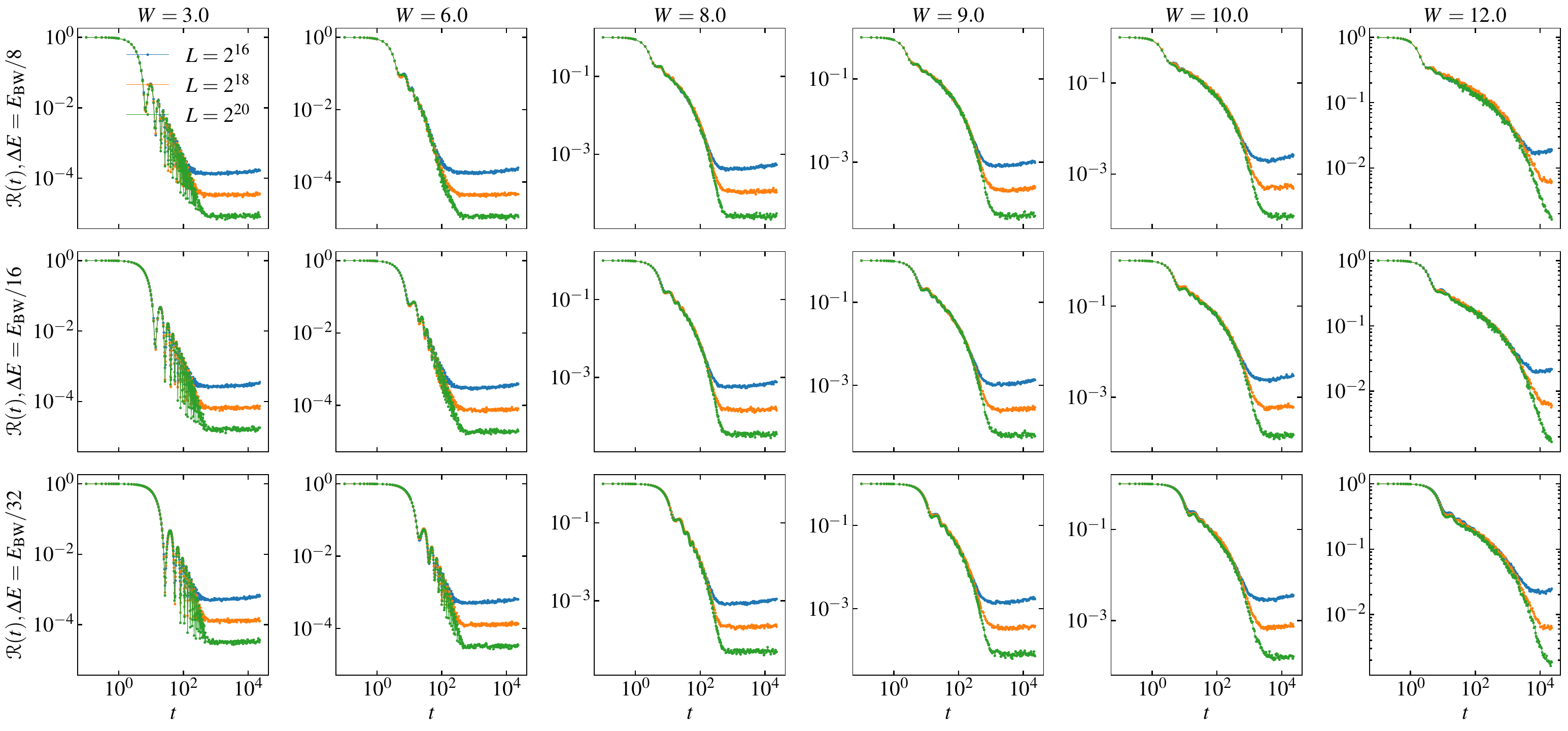}
\caption{The typical value of the return probability $e^{\ovl{\log{\mc{R}}}(t)}$ for several $W$ and for the three largest system sizes $L=2^{16}, 2^{18}, 2^{20}$ and for several energies shells ($|\Delta E|/E_{BW} = 1/8$, $1/16$, $1/32$).}
\label{fig:Typical_rrg}
\end{figure}
\begin{figure}[htb]
\centering
\includegraphics[width=0.75\columnwidth]{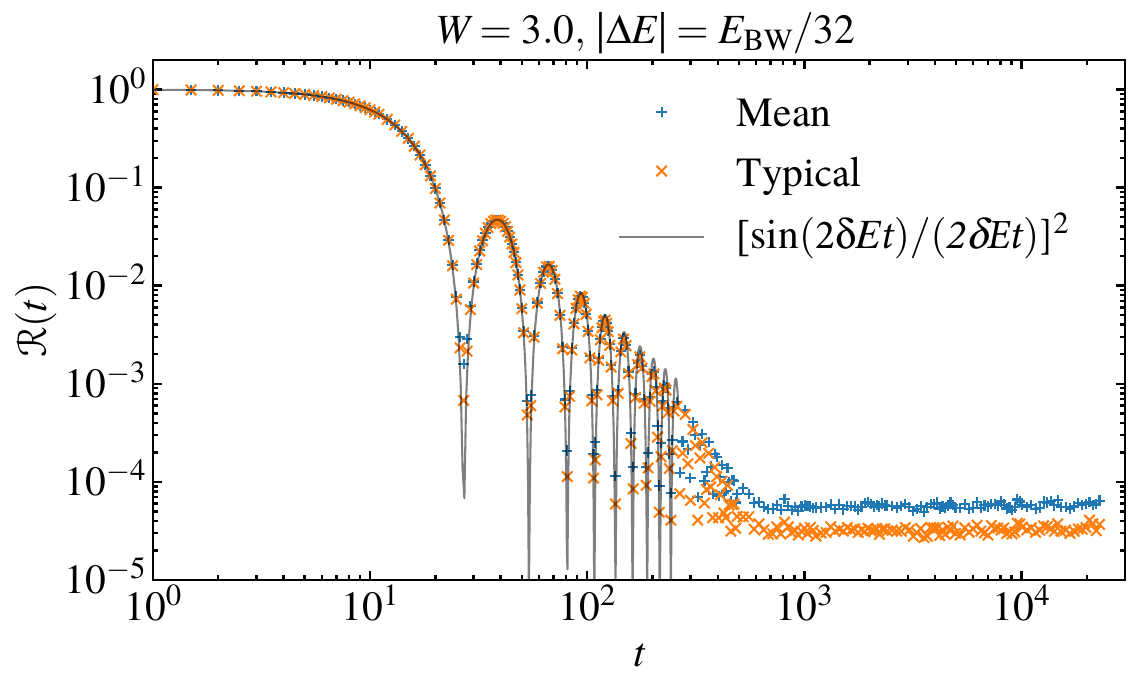}
\caption{The mean and the typical value of the return probability $\mc{R}(t)$ for $W=3$ and $|\Delta E|=2 \delta E = E_{BW}/32$ for the
largest considered system size $L=2^{20}$. $\ovl{\mc{R}}(t)$ and $e^{\ovl{\log{\mc{R}}}(t)}$ decay with time in the same way $\left[\sin(2\delta E t)/2\delta E t \right]^2$.}
\label{fig:Err_rrg}
\end{figure}

Figure~\ref{fig:Err_rrg} shows mean and typical values of $\mathcal{R}(t)$ in the ergodic phase $W=3$ for $L=2^{20}$. As described in the main text in the ergodic phase
there is no distinction between the way of tacking the disorder average (mean or typical value).
Moreover, in the main text, we claim that $\Gamma \ne \Gamma_{{typ}}$ for $0.4 \le W/W_{AT}\le 0.7$.  In what follows, we support our claim doing a different analysis of $\mc{R}(t)$ which does not evolve any fitting procedure.

The idea is to consider separately $\log{ \ovl{\mc{R} }}(t)/\ovl{\log{\mc{R} }}(t)$ and $\log\ovl{\mc{R} }(t)-\ovl{\log{\mc{R} }}(t)$. If a stretched exponent is assumed for the decay of $\mc{R}(t)$
\begin{equation}\label{eq:log_over_log}
 \log{ \ovl{\mc{R} }}(t)/\ovl{\log{\mc{R} }}(t) \sim \frac{\log A - \Gamma t^{\beta} }{\log A_{\text{typ}} - \Gamma_{\text{typ}} t^{\beta} },
\end{equation}
\begin{equation}\label{eq:log_-_log}
\log\ovl{\mc{R} }(t)-\ovl{\log{\mc{R} }}(t)  \sim \log{\frac{A}{A_{\text{typ}}}} - (\Gamma - \Gamma_{\text{typ}}) t^{\beta}.
\end{equation}
In the long time limit $\log{ \ovl{\mc{R} }}(t)/\ovl{\log{\mc{R} }}(t) \rightarrow \frac{\Gamma}{\Gamma_{\text{typ}}}$, nevertheless this regime requires
\begin{equation}
t^{\beta}\gg \frac{\log A_{\text{typ}}}{\Gamma_{\text{typ}}}, \frac{|\Gamma_{\text{typ}}\log A - \Gamma \log A_{\text{typ}}|}{\Gamma \Gamma_{\text{typ}}}.
\end{equation}
Figure~\ref{fig:log_log_rrg1} shows $\log{ \ovl{\mc{R} }}(t)/\ovl{\log{\mc{R} }}(t)$ for
$W=11$, underlining  the plateau ($\log{ \ovl{\mc{R} }}(t)/\ovl{\log{\mc{R} }}(t) \sim \frac{\Gamma}{\Gamma_{\text{typ}}}< 1$).
Furthermore,  $\log \ovl{\mc{R} }(t)-\ovl{\log{\mc{R} }}(t)\sim t^\beta$, what also implies that $\Gamma \ne \Gamma_{\text{typ}}$.

\begin{figure}[htb]
\centering
\includegraphics[width=0.8\columnwidth]{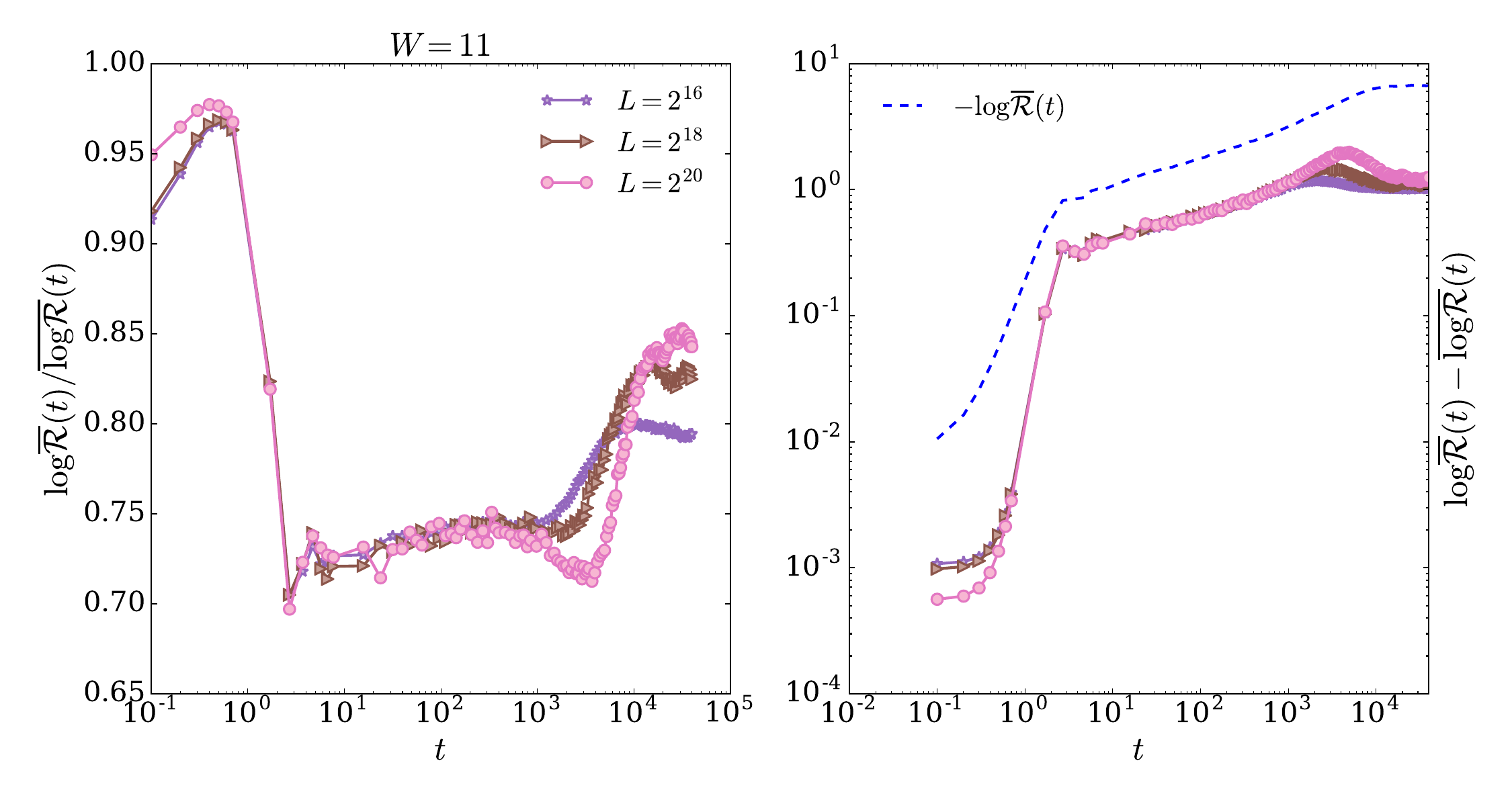}
\caption{The figure shows $\log{ \ovl{\mc{R} }}(t)/\ovl{\log{\mc{R} }}(t)$ and $\log\ovl{\mc{R} }(t)-\ovl{\log{\mc{R} }}(t)$ for $W=11$ and for several system sizes $L$. The blue dashed line in the right panel shows $-\log{ \ovl{\mc{R} }}(t)\sim \Gamma t^\beta$ parallel
to the data \eqref{eq:log_-_log} in log-log scale providing the evidence of the same $\beta$ and unequal $\Gamma\ne\Gamma_{\text typ}$. In these simulation we used $|\Delta E| = 2$.}
\label{fig:log_log_rrg1}
\end{figure}

\subsection{Probability distribution of $\text{IPR}_x$}
\begin{figure}[htb]
\centering
\includegraphics[width=0.8\textwidth]{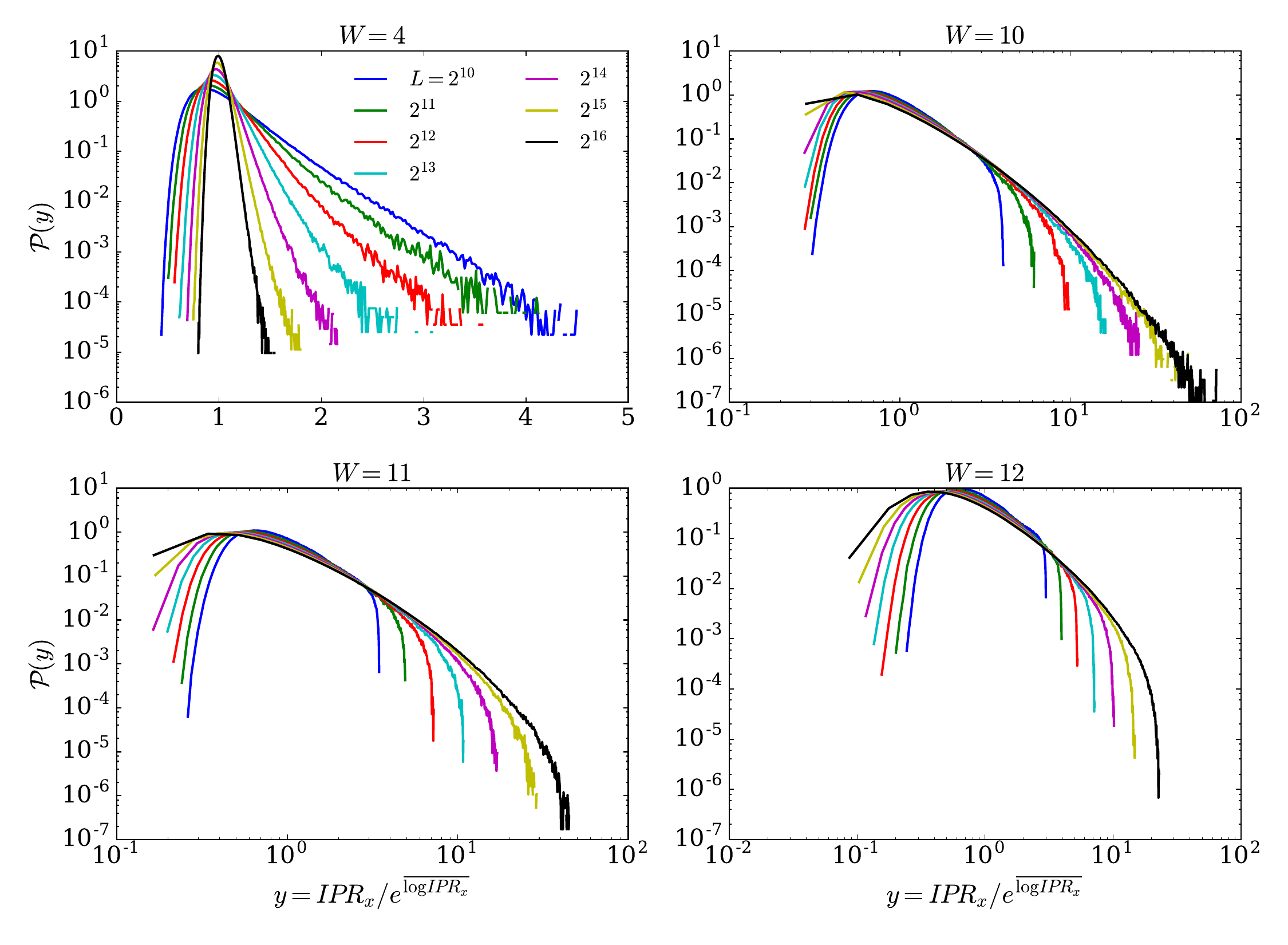}
\caption[Probability distribution for the inverse participation ratio for the random regular graph]
{The figure shows 
the probability distribution of the rescaled variable $y = \text{IPR}_x/e^{\ovl{\log{\text{IPR}_x}}}$.
The panels correspond to different values of $W$. }
\label{fig:Prop_rrg}
\end{figure}
Like in the previous section for PLBM, here we also analyze the full probability distribution of $\text{IPR}_x$ at different disorder strength for
the RRG.
Figure~\ref{fig:Prop_rrg} shows the probability distribution $\mc{P}(y)$ of the rescaled variable $y =
\text{IPR}_x/e^{\ovl{\log{\text{IPR}_x}}}$ for disorder values $W={4, 10, 11, 12}$.
In the so-called ergodic phase $W=2$, the $\mc{P}(y)$ shrinks with $L$ as seen also for the PLBM ergodic phase.
It further strengthens the claims that in this phase $\ovl{\text{IPR}_x}\sim e^{\ovl{\log{\text{IPR}_x}}}$.
For larger values of $W$, the distribution function $\mc{P}(y)$ develops a long tail (algebraically decaying).
This indicates the fact that for these system sizes the $L$-scaling of the mean and typical values of $\text{IPR}_x$ is different. 
Therefore for these values of $W$ the system develops non-ergodic behavior.
%

\end{widetext}
\end{document}